\documentclass{ptephy_v1}
\usepackage{amssymb}
\usepackage{amsmath,amscd}
\usepackage{color}
\usepackage{mathrsfs}

\usepackage{bm}
\usepackage[normalem]{ulem}

\newcommand{\bcen}{\begin{center}}
\newcommand{\ecen}{\end{center}}
\newcommand{\bflr}{\begin{flushright}}
\newcommand{\eflr}{\end{flushright}}
\newcommand{\bfll}{\begin{flushleft}}
\newcommand{\efll}{\end{flushleft}}
\newcommand{\beq}{\begin{equation}}
\newcommand{\eeq}{\end{equation}}
\newcommand{\beqa}{\begin{eqnarray}}
\newcommand{\eeqa}{\end{eqnarray}}
\newcommand{\bite}{\begin{itemize}}
\newcommand{\eite}{\end{itemize}}
\newcommand{\benu}{\begin{enumerate}}
\newcommand{\eenu}{\end{enumerate}}

\newcommand{\cS}{{\mathfrak{S}}}

\newcommand{\cO}{{\cal O}}
\newcommand{\eq}[1]{(\ref{#1})}

\newcommand{\hk}{\hat k}
\newcommand{\hR}{\hat R}

\newcommand{\be}{{\bar \epsilon}}
\newcommand{\bd}{{\bar \delta}}

\newtheorem{theorem}{Theorem}
\newtheorem{definition}{Definition}

\begin{document}

\title{Area bound for surfaces in generic gravitational field}

\author{Keisuke Izumi${}^{1,2}$}
\author{Yoshimune Tomikawa${}^3$}
\author{Tetsuya Shiromizu${}^{2,1}$}
\author{Hirotaka Yoshino${}^4$}

\affil{${}^1$Kobayashi-Maskawa Institute, Nagoya University, Nagoya 464-8602, Japan}
\affil{${}^2$Department of Mathematics, Nagoya University, Nagoya 464-8602, Japan}
\affil{${}^3$Faculty of Economics, Matsuyama University, Matsuyama 790-8578, Japan}
\affil{${}^4$Advanced Mathematical Institute, Osaka City University, Osaka 558-8585, Japan}

\begin{abstract}
  We define an attractive gravity probe surface (AGPS)
  as a compact 2-surface $S_\alpha$ 
with positive mean curvature $k$ satisfying $r^a D_a k / k^2 \ge \alpha$ 
(for a constant $\alpha>-1/2$)
in the local inverse mean curvature flow, where 
$r^a D_a k$ is the derivative of $k$ in the outward unit normal direction.
For asymptotically flat spaces, 
any AGPS is proved to 
satisfy the areal inequality
$A_\alpha \le 4\pi [ ( 3+4\alpha)/(1+2\alpha) ]^2(Gm)^2$, 
where $A_{\alpha}$ is the area of $S_\alpha$ and $m$ is
the Arnowitt-Deser-Misner (ADM) mass.
Equality is realized when the space is isometric
to the $t=$constant hypersurface of the Schwarzschild spacetime
and $S_\alpha$ is an $r=\mathrm{constant}$ surface with $r^a D_a k / k^2 = \alpha$. 
We adapt the two methods of the inverse mean curvature flow and the conformal flow. 
Therefore, our result is
applicable to the case where $S_\alpha$ has multiple components. 
For anti-de Sitter (AdS) spaces, a similar inequality is derived, but
the proof is performed only by using the inverse mean curvature flow. 
We also discuss the cases with asymptotically locally AdS spaces. 
\end{abstract}

\maketitle

\section{Introduction}

A strong gravitational field can make any objects trapped in a region. 
Even photons cannot escape from such a region, which leads to the existence of a black hole. 
A strong gravitational field would be created
by compactly concentrated gravitational sources, 
and thus the trapped region must be small.
The Penrose inequality~\cite{PI} is an argument regarding the area of its boundary; 
the area of an apparent horizon $A_{\rm AH}$ must satisfy
$A_{\rm AH} \le 4 \pi (2Gm)^2$, 
where $m$ is the Arnowitt-Deser-Misner (ADM) mass
and $G$ is Newton's gravitational constant (we use the unit $c=1$). 
The proof for the Penrose inequality has been carried out
for time-symmetric initial data~\cite{J&W,H&I,Bray}. 
To be more precise, those results are applicable for a minimal surface (MS) in an asymptotically flat space with a Ricci scalar
that is non-negative everywhere.
This version of the inequality is called the Riemannian Penrose inequality.

The Riemannian Penrose inequality may seem to provide a
way to test the theory of general relativity. 
However, if the cosmic censorship holds,
an apparent horizon $A_{\rm AH}$ is hidden by an event horizon, 
inside of which we cannot observe, according to the definition of a black hole. 
It is desired to derive an inequality that is 
applicable to a surface outside the event horizon
for the purpose of testing general relativity. 
The authors of this paper have succeeded
in deriving such an inequality~\cite{Shiromizu:2017ego}; we introduced  
the loosely trapped surface (LTS) as a surface 
with positive mean curvature $k>0$ satisfying 
the non-negativity of the derivative of $k$ in the outward unit normal direction 
$r^a D_a k  \ge 0$,
and proved that the LTS satisfies an areal inequality, 
$A_{\rm LTS} \le 4 \pi (3Gm)^2$, where $A_{\rm LTS}$ is the area of the LTS. 
The upper bound is realized if and only if the space is isometric to the time-constant slice of the Schwarzschild spacetime 
and the surface is the photon sphere. 
The photon sphere exists outside the horizon, and thus
an LTS is observable for a distant observer. 
The proof for the areal inequality of the LTS
is achieved with the inverse mean curvature flow
in Ref.~\cite{Shiromizu:2017ego}.

In this paper, we further generalize the Riemannian Penrose inequality
so that it is applicable to a surface in a region
where the gravity is moderately strong
or even weak. 
The surface that we consider in this paper
is a compact 2-surface $S_\alpha$
with positive mean curvature $k$ 
that satisfies $r^a D_a k / k^2 \ge \alpha$ 
(for a constant $\alpha > -1/2$) in the local inverse mean curvature flow,
which we name 
the attractive gravity probe surface (AGPS).%
\footnote{For a Schwarzschild spacetime,
  $r^aD_ak+k^2/2={}^{(3)}R_{ab}r^ar^b=2m/r^3>0$ holds. 
  Then, the condition may reflect the positivity of mass or gravity as an attractive force.
This is the reason why we adopt the name ``AGPS''.} 
We show that an AGPS with $\alpha$ satisfies
$A_{\alpha} \le 4\pi \left[(3+4\alpha)/(1+2\alpha)\right]^2(Gm)^2$, 
where $A_{\alpha}$ is the area of the AGPS. 
For $\alpha \to \infty$, the AGPS becomes the MS
and the Riemannian Penrose inequality is obtained. 
Moreover, for $\alpha = 0$, we have the theorem for
the LTS in Ref.~\cite{Shiromizu:2017ego}. 
Therefore, the inequality for the AGPS is
the generalization of the Riemannian Penrose inequality and 
the areal inequality for the LTS. 
We derive the inequality in two different ways;
one is by using the inverse mean curvature flow \cite{Geroch,J&W,H&I}
and 
the other is by using Bray's conformal flow \cite{Bray}.
Both methods were applied for proving the Riemannian Penrose inequality.
Since Bray's proof is applicable to a surface 
with multiple components, our inequality can be applied to
an AGPS with multiple components as well.

We also show areal inequalities for AGPSs in asymptotically
(locally) anti-de Sitter (AdS)  spaces. 
Since Bray's proof based on the conformal flow is not applicable to
asymptotically (locally) AdS spaces, 
our inequality is only derived by using the inverse mean curvature flow
for these setups. 

The rest of this paper is organized as follows. 
In Sect.~\ref{secIMCF} 
we show the proof by usage of the inverse mean curvature flow  
in the case for asymptotically flat and AdS spaces. 
We also discuss the case for asymptotically locally AdS spaces. 
In Sect.~\ref{secCF}, we show another proof of the areal inequality
for asymptotically flat spaces, by applying Bray's method~\cite{Bray}.  
We will give a summary and discussions in Sect.~\ref{secSum}.  
An example of the matching function used in Sect.~\ref{secCF} is
presented in Sect.~\ref{App}.

\section{Inverse mean curvature flow}\label{secIMCF}

One of the ways to prove the Riemannian Penrose inequality
is applying the monotonicity of the Geroch energy
in the inverse mean curvature flow. 
The idea of the monotonicity was introduced in order to prove the positivity
of the ADM mass by Geroch~\cite{Geroch}.
By applying  Geroch's monotonicity,
the Riemannian Penrose inequality was proved by Wald and Jang~\cite{J&W} 
under the assumption of the existence of a global inverse mean curvature flow. 
This method was extended by Huisken and Ilmanen~\cite{H&I}
to resolve the possible singularity formation in the flow. 
Hence, the monotonicity exists even if
a global inverse mean curvature flow cannot be taken. 
Geroch's monotonicity was generalized
to spaces with negative cosmological constants
by Boucher, Horowitz and Gibbons~\cite{Boucher:1983cv,Gibbons:1998zr}. 

We first review the monotonicity with and without the negative cosmological constant, based on Ref.~\cite{Gibbons:1998zr}, 
in Sect.~\ref{GM}. Then, we show the generalized
areal inequality in Sect.~\ref{IMCFGI}. 
It is applicable also for asymptotically AdS spaces. 
The areal inequality is further generalized
to asymptotically locally AdS spaces,
which is shown in Sect.~\ref{IMCFGI2}.

\subsection{Geroch Monotonicity}\label{GM}

We consider a three-dimensional space $\Sigma$
where the scalar curvature ${}^{(3)}R$ satisfies 
\begin{eqnarray}
{}^{(3)}R -2\Lambda \ge 0 \qquad (\Lambda\le 0). \label{R-2L}
\end{eqnarray}
Let us introduce an inverse mean curvature flow. We first consider the foliation of $\lbrace S_y \rbrace_{y \in {\bf R}}$. 
The lapse function is denoted  by $\varphi$.
Geroch's quasilocal energy of $S_y$ is defined as 
\begin{eqnarray}
E(y):=\frac{A^{1/2}(y)}{64\pi^{3/2}G}\int_{S_y}\left( 2{}^{(2)}R-k^2-\frac{4\Lambda}{3}\right) dA, 
\label{GE}
\end{eqnarray}
where $A(y)$ is the area,
$dA$ is the areal element, and ${}^{(2)}R$
is the scalar curvature of $S_y$. 
Under the condition of the inverse mean curvature flow $k\varphi  =1$,  
the first derivative of $E(y)$ is shown
to satisfy the non-negativity
\begin{equation}
\frac{dE(y)}{dy} \ = \ \frac{A^{1/2}(y)}{64\pi^{3/2}}
\int_{S_y}\Bigl[ 2 \varphi^{-2}({\cal D} \varphi)^2
  +\tilde k_{ab}\tilde k^{ab} + {}^{(3)}R -2 \Lambda \Bigr] dA \ge 0,
\label{E/y}
\end{equation}
where $\tilde k_{ab}$ is the traceless part of the extrinsic curvature $k_{ab}$, {\it i.e.},
\begin{equation}
\tilde k_{ab} := k_{ab} - \frac12 k g_{ab}, 
\end{equation}
with $g_{ab}$ being the metric of $S_y$.
Here, 
we used the assumption that the topology of $S_y$
does not change under the flow,
and thus, $\int {}^{(2)} R dA$ is a constant for all $S_y$ 
because it is the topological invariant. 
Therefore, $E(y)$ is an increasing function,
which leads to $E(\infty)\ge E(y)$.%
\footnote{
  The extension including the singularity resolution
  in the inverse mean curvature flow was performed
  by Huisken and Ilmanen~\cite{H&I}
  under the assumption ${}^{(3)}R \ge 0$. 
  It may be true in the case with the negative cosmological constant
  as pointed out in Ref.~\cite{Gibbons:1998zr}.}
Since 
Geroch's energy at infinity coincides with the ADM mass
in an asymptotically flat spacetime and with 
the Abbott-Deser mass \cite{Abbott:1981} or the Ashtekar-Magnon mass \cite{Ashtekar:1984}
in asymptotically AdS spacetimes,
the inequality implies $m \ge E(y)$. 
If $E(y)$ is expressed with the area of $S_y$, we have an areal inequality
as shown below.

\subsection{Attractive Gravity Probe Surface} \label{IMCFGI}

In an asymptotically flat space without the cosmological constant, 
$E(y)$ for a minimal surface (MS) and for
a loosely trapped surface (LTS)~\cite{Shiromizu:2017ego} is evaluated as
\begin{eqnarray}
E(y)=\dfrac1{2G} \sqrt {\dfrac{A(y)}{4\pi}} \qquad  \mbox{and} \qquad
E(y)\ge \dfrac1{3G} \sqrt {\dfrac{A(y)}{4\pi}},
\end{eqnarray}
respectively. 
Together with the inequality $m \ge E(y)$, they give
the inequalities $A(y) \le 4 \pi (2Gm)^2$ for an MS
and $A(y) \le 4 \pi (3Gm)^2$ for an LTS. 
Let us show similar areal inequalities
by considering a broader class of surfaces, which includes MSs and LTSs.

We consider a surface $S_\alpha$ satisfying $k>0$ and
\begin{eqnarray}
\frac{r^aD_a k}{k^2} \ge \alpha, \label{rDk/k^2}
\end{eqnarray}
where $\alpha$ is a constant satisfying
\begin{equation}
  \alpha > - \frac12,
  \label{alpha-condition}
\end{equation}
and $r^a$ is the outward unit normal vector to $S_\alpha$, namely $r_a= \varphi D_a y$.
We call a surface satisfying the above conditions {\it an attractive gravity probe surface} (AGPS).%
\footnote{
In the next section, we will give a precise definition of AGPS. 
For the proof with the conformal flow, the derivative of $k$ ({\it i.e.}, $r^aD_a k$) must be taken for the local foliation of the inverse mean curvature flow around $S_\alpha$. 
Therefore, this condition is imposed in the definition of AGPS. 
However, it is unnecessary for the proof in this section.
}  
The limit $\alpha \to \infty$ gives the condition for an MS, {\it i.e.} $k=0$,
while the case with $\alpha =0$ is corresponding to
an LTS~\cite{Shiromizu:2017ego}. 
Thus, this surface is a generalization of MSs and LTSs.
Note that in the limit $\alpha\to -1/2$, $S_{\alpha}$ coincides
with the $S^2$ surface at spacelike infinity.
In this sense, the concept of the AGPS can cover from
the surfaces in extremely strong gravity regions to those
in very weak gravity regions
through choosing the value of $\alpha$ appropriately.
We show that its area $A_\alpha$ satisfies an inequality
similar to the Penrose inequality.

On $S_\alpha$, we have a geometrical identity 
\begin{equation}
r^a D_a k \ = \ -\varphi^{-1}{\cal D}^2 \varphi -\frac{1}{2}{}^{(3)}R+\frac{1}{2}{}^{(2)}R  -\frac{1}{2}(k^2+k_{ab}k^{ab}). \end{equation}
Integrating this relation over $S_\alpha$, we have
\begin{eqnarray}
\frac{1}{2}\int_{S_\alpha}{}^{(2)}RdA
&=&
\int_{S_\alpha} \left[r^aD_ak+\varphi^{-2}\left(\cal{D}\varphi\right)^2
  +\frac{1}{2}{}^{(3)}R+\frac{1}{2}\tilde k_{ab}\tilde k^{ab}+\frac{3}{4}k^2 \right] dA
\nonumber\\
&\ge&
\int_{S_\alpha}\left[\left(\frac{3}{4}+ \alpha \right)k^2+\Lambda \right]dA,
\label{Rge}
\end{eqnarray}  
where we used the inequalities of Eqs.~(\ref{R-2L}) and (\ref{rDk/k^2}). 
Note that in the case without a cosmological constant, $\Lambda =0$,
this inequality leads to the positivity of its left-hand side. 
Thus, as a consequence of the Gauss-Bonnet theorem,
the topology of $S_\alpha$ is $S^2$. 
Moreover, for a generic negative cosmological constant, $\Lambda\le 0$,
the Gauss-Bonnet theorem gives us 
\begin{eqnarray}
\int_{S_\alpha}k^2dA\le \left(\frac{3}{4}+ \alpha \right)^{-1} \left[ 2\pi \chi  -\Lambda A_\alpha \right],
\end{eqnarray}
where $A_\alpha$ is the area of $S_\alpha$ and
$\chi$ is the Euler characteristic of $S_\alpha$. 
The Geroch energy $E(y)$ for the surface $S_\alpha$ is estimated as 
\begin{eqnarray}
\left. E(y)\right|_{S_\alpha} &=& 
\frac{A_{\alpha}^{1/2}}{64\pi^{3/2}G}\int_{S_\alpha}\left( 2{}^{(2)}R-k^2-\frac{4\Lambda}{3}\right) dA
\nonumber \\
&\ge&  \frac{1+2\alpha}{3+4\alpha} \frac{\chi}{2G} \left(\frac{A_{\alpha}}{4\pi} \right)^{1/2}  - 
\frac{2\alpha}{3(3+4\alpha)} \frac{\Lambda}{G} \left(\frac{A_{\alpha}}{4\pi} \right)^{3/2}.
\label{m>AA3}
\end{eqnarray}
Under the assumption of the global existence of
the inverse mean curvature flow, the topology of $S_\alpha$ should be $S^2$ 
even in the case with negative cosmological constant, 
because the flow cannot change the topology of all leaves of the
foliation. 
Then, the inequality $m\ge E(y)$ gives us
\begin{eqnarray}
Gm \ge  \frac{1+2\alpha}{3+4\alpha} \left(\frac{A_{\alpha}}{4\pi} \right)^{1/2}   - 
\frac{2\alpha}{3(3+4\alpha)} \Lambda \left(\frac{A_{\alpha}}{4\pi} \right)^{3/2}.
\label{m>AA3}
\end{eqnarray}

Equality in the inequality of Eq.~(\ref{m>AA3}) occurs
if and only if equalities hold in the inequalities
in Eqs.~(\ref{E/y}) and (\ref{Rge}), {\it i.e.},
all ${\cal D}_a \varphi$, $\tilde k_{ab}$ and
${}^{(3)}R-2\Lambda$ vanish on all of $S_y$ and $r^a D_a k /k^2=\alpha$
holds on $S_\alpha$. 
Hence, the metric of $\Sigma$ is 
\begin{eqnarray}
  dl^2 = \left( 1 - \frac{2Gm}{r}
  -\frac{\Lambda}{3} r^2 \right)^{-1} dr^2 + r^2 d \Omega^2, \label{MSAdS}
\end{eqnarray}
which corresponds to the metric of the maximal slice of
an AdS-Schwarzschild spacetime
(or, for $\Lambda=0$, a Schwarzschild spacetime),
and $S_\alpha$ is located at $r=\mathrm{constant}$
on which $r^a D_a k /k^2=\alpha$ is satisfied. 

Since $\Lambda$ is non-positive, the right-hand side of the inequality
of Eq.~(\ref{m>AA3}) is an increasing function with respect to $A_{\alpha}$,
and the maximum occurs when  equality holds. 
Hence, $A_{\alpha}$ is bounded by the area of the $r=\mathrm{constant}$
surface satisfying $r^a D_a k /k^2=\alpha$ on 
the maximal slice of the AdS spacetime 
for $\Lambda<0$,
and the Schwarzschild spacetime for $\Lambda=0$. 

In asymptotically flat spacetimes without
the cosmological constant ({\it i.e.} $\Lambda=0$), the 
inequality of Eq.~(\ref{m>AA3}) can be rewritten as
\begin{eqnarray}
A_\alpha \le 4 \pi \left( \frac{3+4\alpha}{1+2\alpha} G m \right)^2. \label{AB}
\end{eqnarray}
For $\alpha \to \infty$ (the MS) and for $\alpha =0$ (the LTS),
the quantity in the bracket of the above inequality becomes
\begin{eqnarray}
\frac{3+4\alpha}{1+2\alpha} G m  = 
\left\{  \begin{array}{cc}
2Gm& \quad (\alpha \to \infty)\\
3Gm& \quad(\alpha=0)
\end{array}
\right.
.
\end{eqnarray}
Hence, the inequality of Eq.~(\ref{AB}) includes the known results for
the MS and for the LTS.

\subsection{Asymptotically Locally AdS Spacetime} \label{IMCFGI2}

One often considers asymptotically locally AdS spacetimes.
Simple examples are given by the metric 
\begin{eqnarray}
ds^2= -f(r) dt^2 + f^{-1}(r) dr^2 + r^2 d\omega_\kappa^2,
\end{eqnarray}
with
\begin{eqnarray}
f(r) = \kappa - \frac{2Gm}{r} - \frac{\Lambda}{3} r^2,
\label{metricSch}
\end{eqnarray}
where $\kappa=\pm 1$ or $0$. Here, $d\omega_\kappa^2$ denotes 
the metric of a two-dimensional maximally symmetric surface
with the Gauss curvature $\kappa$:
\begin{eqnarray}
d\omega_\kappa^2 = 
\begin{cases}
d \theta^2 + \sin^2 \theta d \phi^2 & (\kappa=+1) \\
d \theta^2 + d \phi^2 &(\kappa=0) \\
d \theta^2 +  \sinh^2 \theta d \phi^2& (\kappa=-1) \quad .
\end{cases} 
\end{eqnarray}
Note that two-dimensional section of this space
can be compactified into a torus
in the case of $\kappa=0$,
and into a multi-torus in the case of $\kappa=-1$,
and such black holes are called topological black holes.
See Ref.~\cite{Aminneborg:1996} for the procedure
for making topological black holes.

Based on these vacuum solutions, it is natural to
consider asymptotically locally AdS spaces whose metrics asymptote to
\begin{eqnarray}
dl^2=  f^{-1}(r) dr^2 + r^2 d\omega_\kappa^2
\end{eqnarray}
at infinity. 
Here, the two-dimensional sections with the metric $r^2d\omega_\kappa^2$
are compactified so that it has genus $g=0$ for $\kappa=+1$,
$g=1$ for $\kappa=0$ and $g=2, 3, \dots$ for $\kappa=-1$. 
Supposing the existence of a global inverse mean curvature flow, 
the topology of $S_\alpha$ satisfying the inequality of
Eq.~(\ref{rDk/k^2}) is the same as that of
the compactified two-dimensional sections. 
We modify the definition of the Geroch energy as
\begin{eqnarray}
E(y):=\frac{A^{1/2}(y)}{(4 \omega_\kappa)^{3/2}G}\int_{S_y}\left( 2{}^{(2)}R-k^2-\frac{4\Lambda}{3}\right) dA, 
\end{eqnarray}
where $\omega_\kappa$ is the area of the compact two-dimensional surface
with the metric $d\omega_\kappa^2$. 
Then, $E(y)$ converges to $m$ in the limit $y\to\infty$, and we have
the inequality
\begin{eqnarray}
Gm \ge  \frac{1+2\alpha}{3+4\alpha} \kappa \left(\frac{A_{\alpha}}{\omega_\kappa} \right)^{1/2}   - 
\frac{2\alpha}{3(3+4\alpha)}\Lambda \left(\frac{A_{\alpha}}{\omega_\kappa} \right)^{3/2},
\end{eqnarray}
where we used $\kappa \omega_\kappa = 2\pi \chi$. 
Here, the area of the unit compact two-dimensional surface $\omega_\kappa$
is given by
\begin{eqnarray}
\omega_\kappa = 
\begin{cases}
4\pi& (\kappa=+1) \\
\mbox{arbitrary} & (\kappa=0) \\
4\pi(g-1) &(\kappa=-1) \quad .
\end{cases}
\end{eqnarray}

\section{Conformal Flow} \label{secCF}

There is another approach for
proving the Riemannian Penrose inequality
based on the conformal flow (Bray's theorem \cite{Bray}). 
It works for the case of a minimal surface with
multiple components, and thus it is more generic than the approach
based on the inverse mean curvature flow. 
In this section, with the aid of Bray's theorem, we prove our inequality of Eq.~(\ref{AB})
to make it applicable to an AGPS with multiple components. 
Since Bray's approach can only be applied to 
asymptotically flat spaces (not asymptotically AdS spaces), 
our areal inequalities for AGPSs are
only applicable to asymptotically flat spaces
as well. 

In Sect. \ref{SecDef}, we give some definitions required in the theorems and proofs.
Our theorem is explicitly shown in Sect. \ref{MT}. 
Remaining subsections are devoted to the proof. 
In Sect. \ref{SoP}, we show the sketch of the proof. 
The main idea is as follows. Our initial data $\Sigma$ has boundaries
$S^{(i)}$ with $i=1,\dots,n$, each of which is supposed to be one component of an AGPS $S_\alpha$.
We glue a manifold ${\bar \Sigma}^{(i)}$ at each $S^{(i)}$
and we suppose that each ${\bar \Sigma}^{(i)}$ has an inner boundary $\mathcal{S}_0^{(i)}$ that is a minimal surface. 
The union of ${\bar \Sigma}^{(i)}$ and that of the $S_0^{(i)}$ are denoted by $\bar \Sigma$ and $\mathcal{S}_0$, respectively. 
$\bar \Sigma$ is constructed so that Bray's method works for $\Sigma \cup \bar \Sigma$. 
There, we see that the proof is completed if the following two
statements hold: 
that the minimal surface ${\cal S}_0$ is the outermost surface, and that the extension of the manifold is smooth.
The former is proved under certain conditions in Sect. \ref{OMMS}. 
With respect to the latter, the extended manifold $\Sigma \cup \bar \Sigma$ 
is generically nonsmooth ($C^0$-class) on the gluing surface $S_\alpha$. 
However, we can show the existence of a sequence of smooth manifolds,
which uniformly converges to the nonsmooth extended manifold. 
Therefore, we will be able to apply Bray's theorem by taking the limit. 
This issue will be discussed in Sect. \ref{SE}.

\subsection{Terminology and Definitions} \label{SecDef}

Let $\Sigma$ be a smooth, asymptotically flat three-manifold with non-negative scalar curvature.  
Here we recall that $\Sigma$ has boundaries, one of which is infinity and 
others are denoted by $S$. 
Everywhere on the boundaries $S$, the mean curvature is supposed to be positive, 
where the unit normal faces  $\Sigma$. 
(We later define this direction of the normal vector of $S$ as ``{\it outward}''.)

Note that if $S$ is smooth, the mean curvature flow from $S$ can always  be taken at least  for a short interval~\cite{H&I,HP}, 
and we call it the local mean curvature flow. 
The explicit definition is shown below. 

\begin{figure}[b]
 \begin{minipage}{0.49\hsize}
  \begin{center}
   \includegraphics[width=70mm]{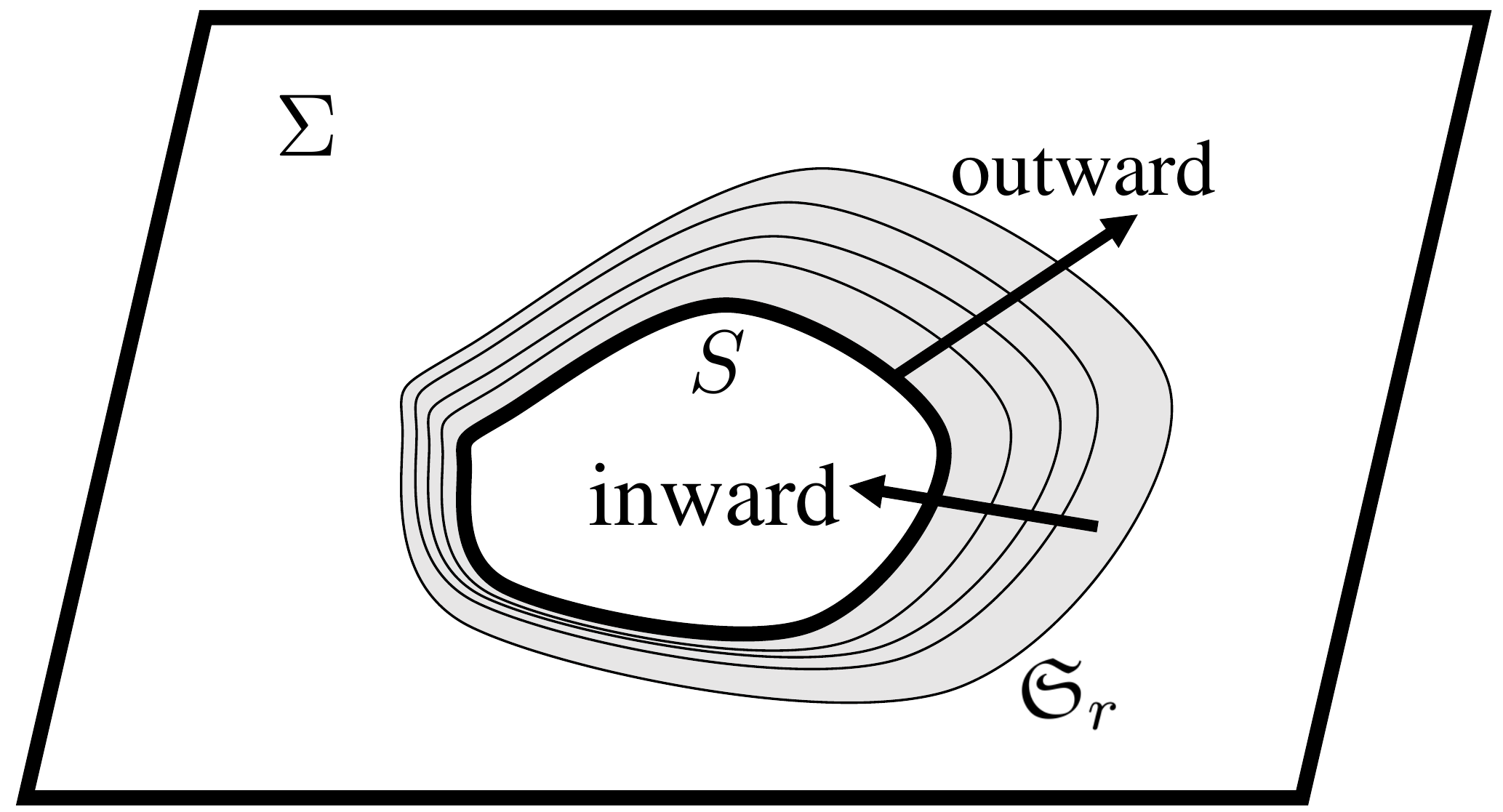}
  \end{center}
  \caption{{\it Outward} and {\it inward} directions of $S$. $S$ (solid curve) is a boundary of $\Sigma$. The {\it outward} and {\it inward} directions are shown. 
The local inverse mean curvature flow with foliations ${\mathfrak{S}}_r$ (thin curves) is taken in {\it exterior} of $S$ (shaded region), if the mean curvature of $S$ with the outward unit normal vector is positive.}
  \label{fig:OI}
 \end{minipage}
 \hspace{0.02\hsize}
 \begin{minipage}{0.49\hsize}
  \begin{center}
   \includegraphics[width=70mm]{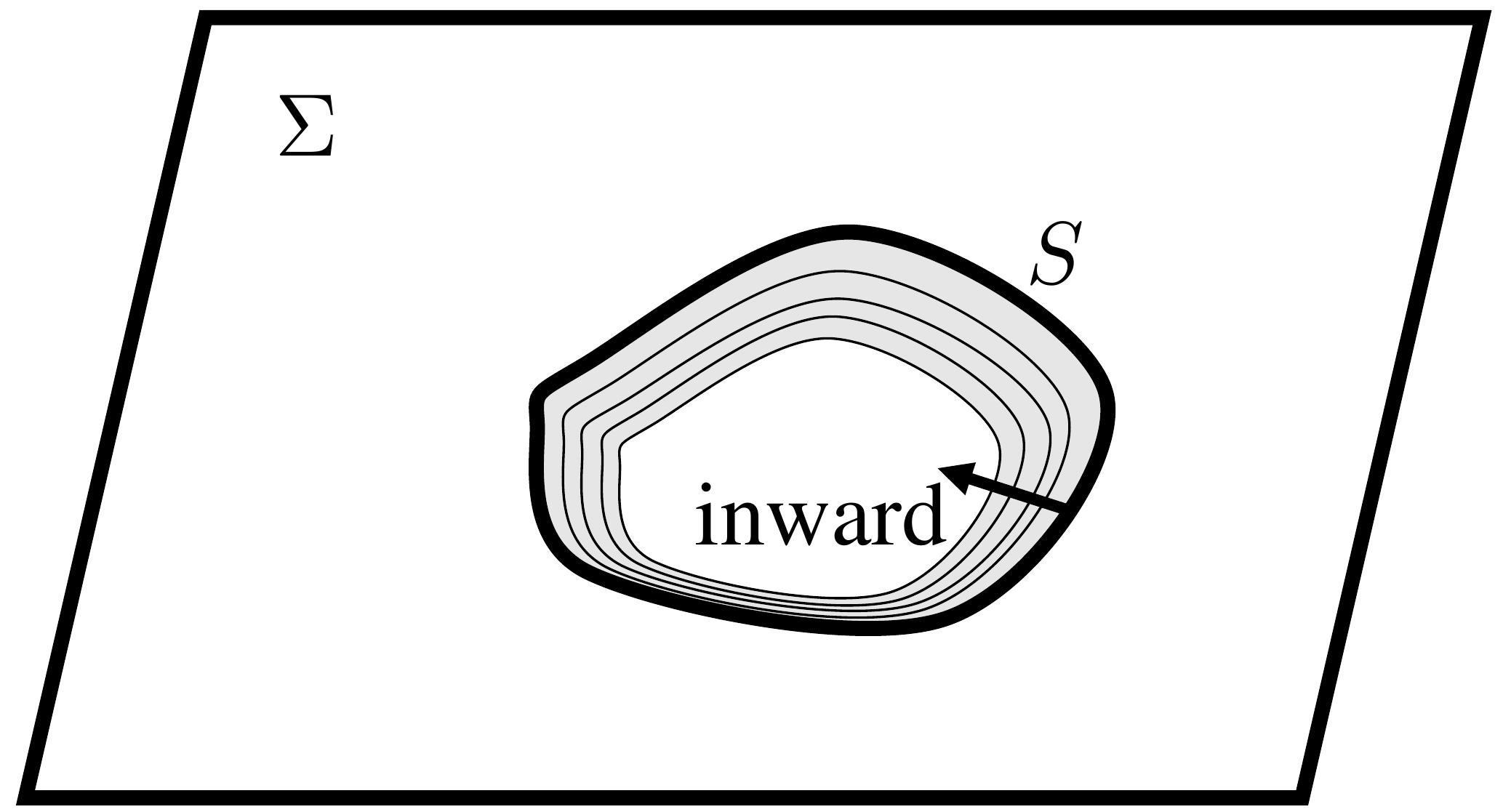}
  \end{center}
  \caption{Inward extension from $S$. $S$ (solid curve)  is a smooth boundary of $\Sigma$. 
  There may exist a smooth extension of $\Sigma$ from $S$, where the local mean curvature flow (with $r<0$) may be taken. The foliations are shown by thin curves. 
  In Theorem~\ref{theorem2}, the exsistence of such an extension is assumed. We name the shaded region {\it interior} of $S$.  }
  \label{fig:Inext}
 \end{minipage}
\end{figure}

\vspace{2mm}
\noindent
%
\begin{definition}
{\bf Local inverse mean curvature flow :}
Let $\Sigma$ to be a smooth three-dimensional manifold. 
$S$, which can be multiple, is a two-dimensional smooth compact manifold embeded in $\Sigma$ (or boundary of $\Sigma$) with 
positive mean curvature.%
\footnote{The positivity of the mean curvature with Eq. (\ref{IMCFeq}) means the positivity of the lapse function for the $r^a$-direction.}
The inverse mean curvature flow is a smooth family of hypersurfaces ${\mathfrak{S}}_r:= x(S,r)$, where 
$x: S \times [0,\varepsilon) \to \Sigma$, satisfying the parabolic evolution equation 
\begin{eqnarray}
\left(\frac{\partial x}{\partial r} \right)^a = \frac{r^a}{k}.
\label{IMCFeq}
\end{eqnarray}
Here, $r^a$ is the unit normal to define the mean curvature $k$ assumed to be positive everywhere on ${\mathfrak{S}}_r$, 
and the initial condition is set to be $S = x(S,0)$.
Since Eq.\eq{IMCFeq} is parabolic, a solution exists for a short interval of $r$, $0\le r < \varepsilon$ \cite{H&I,HP}. 
We call this the  ``{\it local mean curvature flow}'' from $S$.
\end{definition}
%
%

\vspace{2mm}

Here we remind the reader that the Penrose inequality is a theorem for the {\it outermost} minimal surface. 
This means that no other minimal surface exists in its {\it exterior region}. 
Similar conditions are required for our theorem, and then we define the term ``{\it enclose}'' as follows. 
Let $\sigma$ be a compact two-dimensional closed surface (which can be multiple) in a three-dimensional manifold $\Sigma$ with boundary components as well as a single asymptotically flat end. 
Let ${\cal R}$ be a region which consists of all points $p \in \Sigma\backslash\sigma$ path-connected with the end in $\Sigma\backslash\sigma$. 
We call ${\cal R}$ the ``{\it exterior}'' region of $\sigma$.
We also define the ``{\it interior}'' region of $\sigma$ as $\bar{\cal R} = \Sigma\backslash \left({\cal R} \cup \sigma \right)$. 
($\bar{\cal R}$ may be empty.)
Then if $\tilde {\cal R} \subset  \bar{\cal R} \cup \sigma $, 
we say that $\sigma$ encloses $\tilde {\cal R}$ (or $\tilde {\cal R}$ is enclosed by $\sigma$).


In the manifold $\Sigma$ that we will consider, 
the boundary $S$ has positive mean curvature with the unit normal vector $r^a$ directing into $\Sigma$, {\it i.e.}, into the {\it exterior} of $S$. 
The local inverse mean curvature flow is taken in the exterior of $S$.
If a direction corresponds to the increase (decrease) in $r$ of Eq.~\eq{IMCFeq}, 
it is called ``{\it outward}'' (``{\it inward}''), {\it i.e.}, 
$r^a$ is the {\it outward} normal vector.
We sometimes consider an extension of $\Sigma$ from $S$, 
which is also called the {\it interior} of $S$.


The areal inequality will be
proved  for attractive gravity probe surfaces (AGPSs) introduced in Sect.~\ref{IMCFGI}. 
To make our theorem precise, we explicitly give the definition of
the AGPS here.

\vspace{2mm}

\noindent
%
\begin{definition}{\bf Attractive gravity probe surfaces (AGPSs) :\ }
  Suppose $\Sigma$ to be a smooth three-dimensional manifold with
  a positive definite metric $g$.
A smooth compact surface $S_\alpha$ in $\Sigma$ is an attractive gravity probe surface (AGPS) with a parameter $\alpha$ ($\alpha>-1/2$)
if the following conditions are satisfied everywhere on $S_\alpha$: 
\begin{description}
\setlength{\leftskip}{5mm}
 \item[(i)]\ The mean curvature $k$ is positive.
 \item[(ii)]\ With the local mean curvature flow in the neighborhood of $S_\alpha$,  
\begin{eqnarray}
r^a D_a k \ge \alpha k^2 \label{rDkineq}
\end{eqnarray}
is satisfied, where $r^a$ is the outward unit normal vector to $S_\alpha$ used for the definition of $k$, and $D_a$ is the covariant derivative with respect to the metric $g$.
\end{description}
\label{definition-AGPS}
\end{definition}
%
%
\vspace{2mm}

\subsection{Main Theorem} \label{MT}
Now, we present our main theorem:

%
\begin{theorem}
Let $\Sigma$ be an asymptotically flat, three-dimensional, smooth manifold with non-negative Ricci scalar. 
The boundaries of $\Sigma$ are composed of an asymptotically flat end and
an AGPS $S_\alpha$, which can have multiple components,
with a parameter $\alpha$ in Eq.~\eq{rDkineq}. 
Here, the unit normal $r^a$ used to define the mean curvature is taken to be outward of $S_\alpha$.
Suppose that
there exists a positive number $\varepsilon$ such that, 
for any leaf ${\mathfrak{S}}_r = x(S_\alpha,r)$ of the foliation  in the local mean curvature flow from $S_\alpha$, $x: S_\alpha \times [0,\varepsilon) \to \Sigma$;
no minimal surface satisfying either one of the following conditions exists:
\begin{description}
\setlength{\leftskip}{5mm}
 \item[(i)] \ It encloses (at least) one component of ${\mathfrak{S}}_r$. 
 \item[(ii)] \  It has boundaries on ${\mathfrak{S}}_r$ and
   its area is less than $4\pi (2Gm)^2$.
\end{description}
Then, the area of $S_\alpha$ has an upper bound; 
\begin{eqnarray}
A_{\alpha} \le 4\pi \left( \frac{3+4\alpha}{1+2\alpha} Gm\right)^2,
\end{eqnarray}
where $m$ is the ADM mass of the manifold,
$G$ is Newton's gravitational constant.
Equality holds if and only if $\Sigma$ is a time-symmetric hypersurface
of a Schwarzschild spacetime and 
$S_\alpha$ is a spherically symmetric surface with  $r^a D_a k = \alpha k^2$. 
\label{theorem}
\end{theorem}
%
%
\vspace{2mm}

In the smooth extension that will be discussed in Sect.~\ref{SE}, we will extend $\Sigma$ from $S_\alpha$ with a deformation in a neighborhood $U (\subset \Sigma)$ of $S_\alpha$, 
and there, the non-existence conditions $(i)$ and $(ii)$ for minimal surfaces are required for the outer boundary of the deformed region, which is not $S_\alpha$ but a leaf of the foliation $\cS_r$ (see Fig.\ref{fig:OI}). 
This is because, in Theorem~\ref{theorem},
the existence of the smooth extension from $S_\alpha$ to the inward
cannot be assumed in general, 
and thus, the deformation should occur in the exterior of $S_\alpha$.
In physical situations, however,
it is natural to suppose that $S_\alpha$ is embedded in a three-dimensional space $\Sigma_p$, 
where  $\Sigma_p$ has not only the exterior regions of $S_\alpha$ but also its interior regions. 
Here, $\Sigma$ is taken as a subset of $\Sigma_p$, whose boundaries are composed of $S_\alpha$ and the flat end. 
This provides us a concrete smooth extension of $\Sigma$ from $S_\alpha$ inwards in $\Sigma_p$.  
Then, the local mean curvature flow can be taken in the interior of $S_\alpha$ and 
the deformation can take place in the interior (see Fig.~\ref{fig:Inext}).
As will be discussed in detail at the end of Sect.\ref{SE},  the outer boundary of the deformed region can be taken to be $S_\alpha$ instead of ${\mathfrak{S}}_r$ (compare between Figs.~\ref{fig:OI} and \ref{fig:Inext}), 
and thus, the non-existence conditions $(i)$ and $(ii)$ are required to be assumed only for $S_\alpha$.
This gives another version of Theorem \ref{theorem}.
%
\begin{theorem}
Let $\Sigma$ be an asymptotically flat, three-dimensional, smooth  manifold with non-negative Ricci scalar. 
The boundaries of $\Sigma$ are composed of an asymptotically flat end and
an AGPS $S_\alpha$, which can have multiple components,
with a parameter $\alpha$. 
Here, the unit normal $r^a$ to define the mean curvature is taken to be outward of $S_\alpha$.
Suppose that there exists a finite-distance smooth extension of $\Sigma$ from $S_\alpha$ to the interior of $S_\alpha$
satisfying the non-negativity of the Ricci scalar, 
such that the local inverse curvature flow from $S_\alpha$ can be taken in the interior region, 
and that
no minimal surface satisfying either one of the following conditions exists:
\begin{description}
\setlength{\leftskip}{5mm}
 \item[(i)] \ It encloses (at least) one component of $S_\alpha$. 
 \item[(ii)] \  It has boundaries on $S_\alpha$ and
   its area is less than $4\pi (2Gm)^2$.
\end{description}
Then, the area of $S_\alpha$ has an upper bound; 
\begin{eqnarray}
A_{\alpha} \le 4\pi \left( \frac{3+4\alpha}{1+2\alpha} Gm\right)^2,
\end{eqnarray}
where $m$ is the ADM mass of the manifold and
$G$ is Newton's gravitational constant.
Equality holds if and only if $\Sigma$ is a time-symmetric hypersurface
of a Schwarzschild spacetime and 
$S_\alpha$ is a spherically symmetric surface with  $r^a D_a k = \alpha k^2$. 
\label{theorem2}
\end{theorem}
%
\vspace{2mm}
\noindent
The proof of the theorems is given in the following subsections.

\subsection{Sketch of Proof}\label{SoP}
The proofs for Theorems~\ref{theorem} and \ref{theorem2} are almost the same except for the treatment of an extension of $\Sigma$. 
Therefore, the sketch here is common. 
We will basically present the proof for Theorem~\ref{theorem} first, 
and then give a comment on Theorem~\ref{theorem2} at the end of Sect.~\ref{SE}.

The  local inverse mean curvature flow gives a metric
\begin{eqnarray}
ds^2 = \varphi^2 d r^2 + g_{ab} dx^a dx^b, \label{wohatm}
\end{eqnarray}
where each $r$-constant surface is a leaf of the foliation $\cS_r$ in the local inverse mean curvature flow with $\varphi k = 1$. 
Here, we take $r$ such that $r=0$ on $S_\alpha$ and $r$ increases in
the outward direction. Since $\varphi$ is supposed to have the dimension of length,
the coordinate $r$ is a non-dimensional quantity.

\begin{figure}[tb]
  \begin{center}
    \includegraphics[width=10.0cm,bb=0 0 960 482]{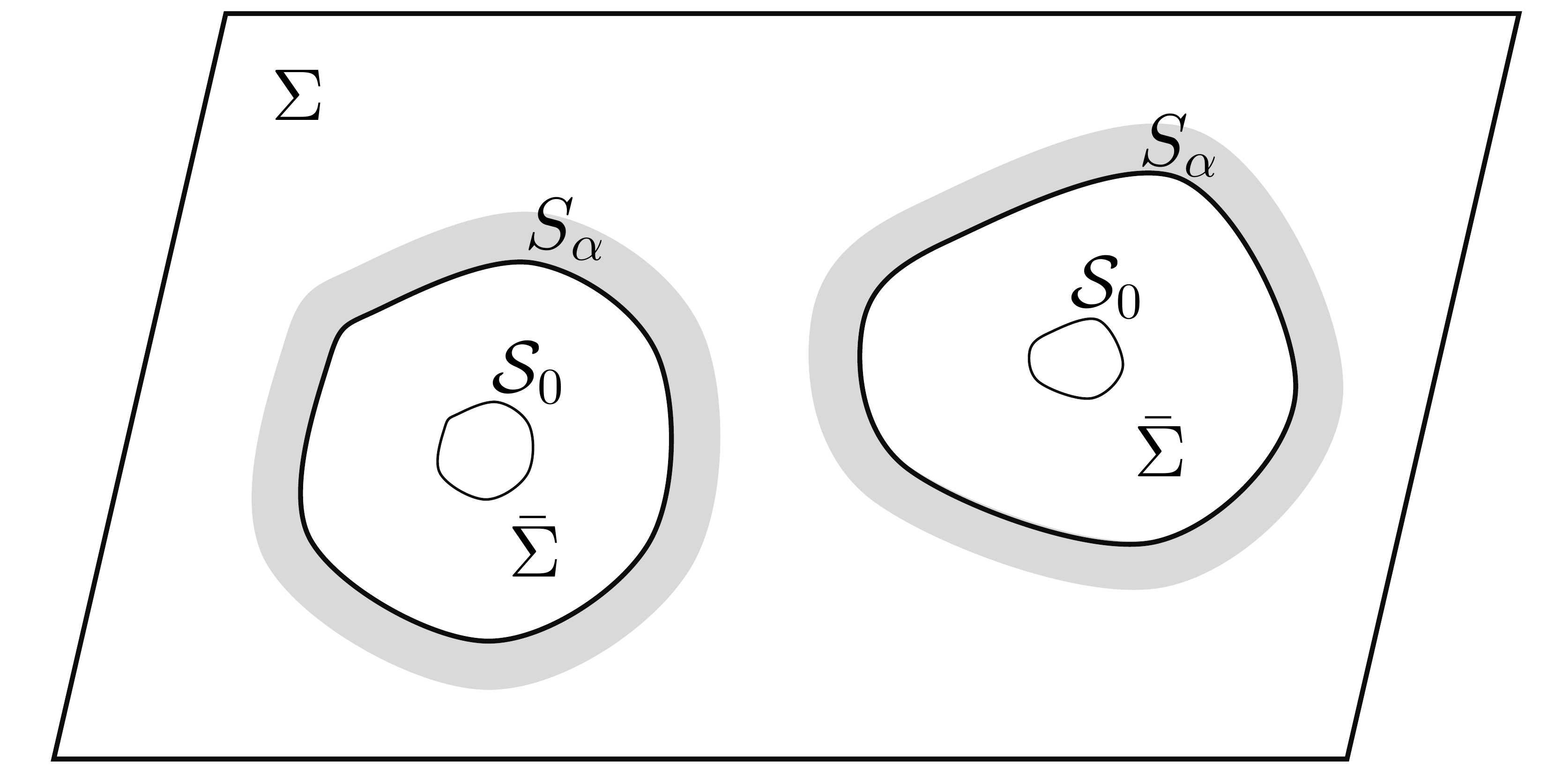}
  \end{center}
  \caption{Extension of the manifold: $\Sigma$ is an asymptotically flat space with a boundary $S_\alpha$, which is an AGPS with the parameter $\alpha$
    (with multiple components, in general). 
      The shaded region is the neighborhood of $S_\alpha$ in $\Sigma$, where the local inverse mean curvature flow of Eq.~\eq{wohatm} is taken. 
      On $S_\alpha$, the manifold $\bar \Sigma$ with the metric of Eq.~\eq{barm} is glued to $\Sigma$.   
      In $\bar \Sigma$, a minimal surface ${\cal S}_0$ exists at $r=-r_0+2\log2 ~(<0)$ in our current setup. }
   \label{Figcon}
\end{figure}

On (each component of) $S_\alpha$,
we attach a manifold $\bar \Sigma$
with a metric in the range $r\le0$ (see Fig.~\ref{Figcon}),
\begin{eqnarray}
d\bar s^2 
=  \frac{ 1-2\exp \left( -\frac{r_0}{2} \right)}{1-2\exp \left( -\frac{r+r_0}{2} \right)}\exp \left( r \right) \varphi_0^2 (x^a) dr^2 
+\exp \left( r \right) g_{0,ab} (x^a) dx^a dx^b,
\label{barm}
\end{eqnarray}
where  $g_{0,ab}:= g_{ab}|_{r=0}$ in the metric of Eq.~(\ref{wohatm})
and $\varphi_0 = \varphi|_{r=0}$. 
Each $r$-constant surface is umbilical, {\it i.e.}, the extrinsic curvature is 
\begin{eqnarray}
&& \bar k_{ab} = \frac12 \bar k \bar g_{ab}. 
\end{eqnarray}
Hereafter, quantities with a bar indicate that they are associated
with the metric of Eq.~\eq{barm}.
The mean curvature and its derivative are 
\begin{eqnarray}
&& \bar k = \varphi_0^{-1} \sqrt{\frac{1-2\exp \left( -\frac{r+r_0}{2} \right)}{ 1-2\exp \left( -\frac{r_0}{2} \right)}\exp \left(- r \right)}, \\
&&\bar{r}^a \bar{D}_a \bar{k} = -\frac12 \bar k^2 \frac{1-3 \exp \left( - \frac{r+r_0}{2} \right) }{1-2 \exp \left( - \frac{r+r_0}{2} \right) }. 
\end{eqnarray}
Note that the lapse function in the metric of Eq.~\eq{barm} has been tuned so that $\bar k|_{S_\alpha} = k|_{S_\alpha}$ is satisfied. 
Meanwhile, $r_0$ is tuned so that 
\begin{eqnarray}
\bar{r}^a \bar{D}_a \bar{k}|_{S_\alpha} =  \alpha \bar k^2|_{S_\alpha}
\end{eqnarray}
holds, which is equivalent to
\begin{eqnarray}
\exp \left(\dfrac{r_0}{2}\right) = \frac{3+ 4\alpha}{1+ 2\alpha}.
\label{r0}
\end{eqnarray}
The manifold $\bar\Sigma$ is continuously glued to $\Sigma$ at $r=0$ because the induced metrics on the gluing  
surface $S_\alpha$ $(r=0)$ are the same. 
However, the metric is generally $C^0$-class there
because $S_{\alpha}$ may not be umbilical, {\it i.e.} 
the extrinsic curvature $k_{ab}$ may have non-vanishing traceless part.

The three-dimensional Ricci scalar of $\bar\Sigma$ is expressed with the two-dimensional quantities as
\begin{eqnarray}
{}^{(3)}\bar R =  {}^{(2)}\bar R - 2 \bar \varphi^{-1} \bar {\cal D}^2 \bar \varphi - 2 \bar{r}^a \bar{D}_a \bar{k} - \frac32 \bar k^2,
\label{SPbarR}
\end{eqnarray} 
where $\bar \varphi$ is the lapse function in the metric of Eq.~\eq{barm},
{\it i.e.}, 
\begin{eqnarray}
\bar \varphi =  \varphi_0 (x^a) \sqrt{ \frac{ 1-2\exp \left( -\frac{r_0}{2} \right)}{1-2\exp \left( -\frac{r+r_0}{2} \right)}\exp \left( r \right) }.
\end{eqnarray} 
Each term in the right-hand side of
Eq.~(\ref{SPbarR}) can be explicitly calculated as
\begin{eqnarray}
&& {}^{(2)}\bar R = \exp (-r) {}^{(2)} R_0,  \qquad   
 \bar \varphi^{-1} \bar {\cal D}^2 \bar \varphi = \exp (-r)  \varphi_0^{-1}  {\cal D}^2  \varphi_0, 
\nonumber \\
&&
   2 \bar{r}^a \bar{D}_a \bar{k} + \frac32 \bar k^2 = \exp (-r) \left(2\alpha +\frac32 \right)\varphi_0^{-2}.
\end{eqnarray} 
Therefore, we have
\begin{eqnarray}
{}^{(3)}\bar R &=&  \exp (-r)\left\{ {}^{(2)} R_0 - 2 \varphi_0^{-1}  {\cal D}^2  \varphi_0 -  \left( 2\alpha +\frac32  \right)\varphi_0^{-2}\right\} \nonumber \\
&=& \exp (-r) {}^{(3)}\bar R_0,
\end{eqnarray} 
where ${}^{(3)}\bar R_0$ is defined by 
\begin{eqnarray}
 {}^{(3)}\bar R_0 :=  \lim_{r \nearrow 0} {}^{(3)}\bar R.
\end{eqnarray} 
Since $\varphi_0^{-2}=k^2 |_{r=0}$ holds in the inverse mean curvature flow
and $r^a D_a k |_{r=0} \ge \alpha k^2|_{r=0}$ holds from the definition
of the AGPS, 
${}^{(3)}\bar R_0$ turns out to be non-negative, 
\begin{eqnarray}
  {}^{(3)}\bar R_0
 & \ge& {}^{(2)} R_0 - 2 \varphi_0^{-1}  {\cal D}^2  \varphi_0 -   \left. 2 r^aD_a k\right|_{r=0} 
 -\left. k^2\right|_{r=0} -\left. k_{ab}^2 \right|_{r=0}  \nonumber \\
&=& {}^{(3)} R_0 \ge 0, 
\end{eqnarray} 
where ${}^{(3)} R_0$ is the Ricci scalar of $\Sigma$ on $S_\alpha$ ($r=0$), {\it i.e.}, 
$ {}^{(3)} R_0 :=  R |_{r=0}$.
Thus, the three-dimensional Ricci scalar of
the glued manifold $\Sigma \cup \bar\Sigma$ is non-negative everywhere. 
Hence, it is expected that Brays' proof to the Riemannian
Penrose inequality~\cite{Bray} by the conformal flow would be applicable
for $\Sigma \cup \bar \Sigma$. 
However, it requires the smoothness of the manifold
as an assumption, while our manifold 
$\Sigma \cup \bar\Sigma$ is generally $C^0$-class, not $C^\infty$-class at the gluing surface.
This problem will be fixed later (see Sect.~\ref{SE});
in the rest of the present subsection,
we assume that Bray's theorem is applicable
to the manifold $\Sigma \cup \bar\Sigma$.

The manifold $\bar\Sigma$ has a minimal surface at $r= -r_0 + 2\log 2$ 
(the value of $-r_0 + 2\log 2$ is always negative
because of Eq.~\eq{r0} and $\alpha> -1/2$). 
The minimal surface is denoted by ${\cal S}_0$. 
From the metric of Eq.~\eq{barm},
we find the direct relation between the area of ${\cal S}_0$ and
that of $S_\alpha$ as
\begin{eqnarray}
A_0 &=& \exp (-r_0 + 2\log 2) A_\alpha \nonumber \\
&=& \left[\frac{2(1+2\alpha)}{3+4\alpha}\right]^2 A_\alpha,
\label{AalphaA0}
\end{eqnarray} 
where $A_0$ is the area of ${\cal S}_0$. 
Supposing ${\cal S}_0$ to be the outermost minimal surface,
the required conditions for which will be discussed in Sect.\ref{OMMS}, 
the Riemannian Penrose inequality~\cite{Bray} gives 
\begin{eqnarray}
A_0 \le 4\pi (2Gm)^2.
\label{PE}
\end{eqnarray} 
Combining this with Eq.~\eq{AalphaA0}, we have 
\begin{eqnarray}
A_\alpha \le 4\pi \left(\frac{3+4\alpha}{1+2\alpha}Gm\right)^2.
\label{MPE}
\end{eqnarray} 

When equality holds in the inequality of Eq.~\eq{MPE},
that for the Riemannian Penrose inequality of Eq.~\eq{PE} holds
as well. 
Then, the manifold has the metric of the time-symmetric slice of
a Schwarzschild spacetime. 
Hence, each of the manifolds $\Sigma$
and $\bar{\Sigma}$ is isometric to a part of it, and
in particular, the surface $r=-r_0+2\log 2$ is a spherically symmetric
minimal surface. 
Hence, because of the construction of the manifold $\bar{\Sigma}$
whose metric is given in Eq.~\eqref{barm},  
$S_\alpha$ should also be spherically symmetric
and must satisfy $r^a D_a k = \alpha k^2$.
As a result, equality in the inequality of Eq.~\eq{MPE}
holds if and only if 
$\Sigma$ is the time-symmetric hypersurface of a Schwarzschild spacetime and 
$S_\alpha$ is the spherically symmetric surface with  $r^a D_a k = \alpha k^2$.

As stated before, the remaining task is to show that the manifold $\Sigma \cup \bar \Sigma$ 
can be made smooth at the gluing surface $S_\alpha$ and ${\cal S}_0$ is the outermost minimal surface. 
We will discuss the latter in Sect.~\ref{OMMS}, and then,
prove the former in Sect.~\ref{SE}.

\subsection{Outermost Minimal Surface}\label{OMMS}

In the previous subsection, we used the assumption
that the minimal surface ${\cal S}_0$ at $r=-r_0+2\log 2$ 
in the extended manifold $\bar\Sigma$ is outermost.
Our Theorems
are carefully stated to guarantee this assumption,
as we explain below. 

For simplicity, the study in this section is given for the non-smooth manifold that we constructed in Sect.~\ref{SoP}, and
the non-existence conditions $(i)$ and $(ii)$ are assumed for $S_\alpha$, not for each foliation $\cS_r$. 
The discussion here must be extended for all smooth manifolds constructed in Sect.~\ref{SE}, 
and we will come back to this point later.

%

A part of  the minimal surface ${\cal S}_0$ at $r=-r_0+2\log 2$ is not outermost
if an outermost minimal surface exists in $r>0$, in $r<0$, or across $r=0$.
We examine these three possible cases, one by one.
The first possibility is prohibited by the assumption
of the theorem that any part of $S_\alpha$ is not enclosed by a minimal surface. 

The second case is impossible, {\it i.e.}, 
we cannot take minimal surfaces in the extended manifold $\bar\Sigma$ not touching with the boundary of $\bar\Sigma$ except the minimal surface $r=-r_0+2\log 2$. 
To show this using proof by contradiction, let us assume that the second case is possible. 
Such a surface has maximum $r$ at some points. 
Then, we can show that the mean curvature $k$ there becomes positive ({\it i.e.}, non-zero), and thus, it is not a minimal surface, which leads to a contradiction.

Let us begin with a generic discussion of geometry. Suppose we have a codimension one surface $\sigma$ in a manifold 
and we take the Gaussian normal coordinates in the neighborhood of $\sigma$,
\begin{eqnarray}
dl^2 = dy^2 + h_{ij} dx^i dx^j, 
\end{eqnarray} 
where the surface $\sigma$ is supposed to exist at $y=0$ and its unit normal vector is
\begin{eqnarray}
n_\mu = \partial_\mu y.
\end{eqnarray} 
We take another surface $\tilde \sigma$ characterized with $\tilde y = 0$ and
\begin{eqnarray}
\tilde y = y - \delta y(x^i).
\end{eqnarray} 
Suppose  that this surface has maximum $y$ at $x^i=0$ and its value is $y=0$. 
Hence,  $\tilde \sigma$ is tangent to $\sigma$ at $x^i=0$ and the Hessian matrix of $\delta y(x^i)$ is negative-semidefinite there, {\it i.e.} we have
\begin{eqnarray}
\left. \partial_\mu \delta y  \right|_{x^i=0}=0, \qquad  \left. h^{ij}\partial_i \partial_j \delta y \right|_{x^i=0} \le 0. 
\label{Hess}
\end{eqnarray} 
The unit normal vector of $\tilde \sigma$ directing to the same side as $n^\mu$ is 
\begin{eqnarray}
\tilde n_\mu = \alpha \partial_\mu \tilde y.
\end{eqnarray} 
with
\begin{eqnarray}
\alpha = \bigl(1+ h^{ij} (\partial_i \delta y) (\partial_j \delta y) \bigr)^{-\frac12}.
\end{eqnarray} 
The induced metric of $\tilde \sigma$ is
\begin{eqnarray}
\tilde h_{\mu\nu} = g_{\mu\nu}- \tilde n_\mu \tilde n_\nu,
\end{eqnarray} 
and $(\tilde h_{\mu\nu}, \tilde n_\mu)$ coincides with $(h_{\mu\nu},  n_\mu)$ at $x^i=0$. 
The mean curvature of $\tilde\sigma$, $\tilde k$, is calculated as 
\begin{eqnarray}
\tilde k := g^{\mu\nu} \nabla_\mu \tilde n_\nu 
= g^{\mu\nu} (\partial_\mu \alpha)   \left( n_\nu -\partial_\nu \delta y\right)
+  \alpha g^{\mu\nu} \left( \nabla_\mu n_\nu - \partial_\mu \partial_\nu \delta y  -  \Gamma^\alpha_{\mu\nu} \partial_\alpha \delta y \right).
\end{eqnarray}
Since we have 
\begin{eqnarray}
&&\alpha|_{x^i=0}=1, \qquad \left. g^{\mu\nu} \partial_\mu \partial_\nu \delta y \right|_{x^i=0} = h^{ij} \partial_i \partial_j \delta y ,\nonumber \\
&& \left. \partial_\mu \alpha\right|_{x^i=0} = \left. - \frac12 \alpha^3 \left[2h^{ij}(\partial_i \delta y) (\partial_\mu \partial_j \delta y) +
(\partial_\mu h^{ij})(\partial_i \delta y) ( \partial_j \delta y)\right]\right|_{x^i=0} =0,
\end{eqnarray}
the mean curvature $\tilde k$ is related to $k$ as  
\begin{eqnarray}
\tilde k |_{x^i=0} =  (k - h^{ij} \partial_i \partial_j \delta y ) |_{x^i=0} \ge k|_{x^i=0}, 
\label{tikk}
\end{eqnarray}
where we used Eq.~\eq{Hess}. 

We apply Eq.~\eq{tikk} to surfaces on $\bar\Sigma$ as follows. 
As the surface $\tilde\sigma$ and the point $x^i=0$, we adopt the minimal surface $\sigma_m$ other than $r=-r_0+2\log 2$ which is assumed to exist and the point $p_M$ where the maximum of $r(=r_M)$ occurs on the minimal surface,   respectively. 
As the surface $\sigma$, we adopt the $r$-constant surface $r=r_M$. 
Then, from Eq.~\eq{tikk}, at $p_M$ the mean curvature $k_{m}$ of $\sigma_m$ satisfies
\begin{eqnarray}
k_{m} \ge k_M, 
\end{eqnarray}
where $k_M$ is the mean curvature of the $r=r_M$ surface. 
Since $r_M$ is larger than $-r_0+2\log 2$, the mean curvature $k_M$ is positive, and thus   
$k_{m}$ should be positive at $p_M$. 
This is inconsistent with the assumption that the surface $\sigma_m$ is minimal. 
Therefore, there exists no minimal surface in $\bar\Sigma$ that is not touching with the boundary of $\bar\Sigma$ other than $r=-r_0+2\log 2$.

\begin{figure}[tb]
  \begin{center}
    \includegraphics[width=15.0cm,bb=0 0 875 242]{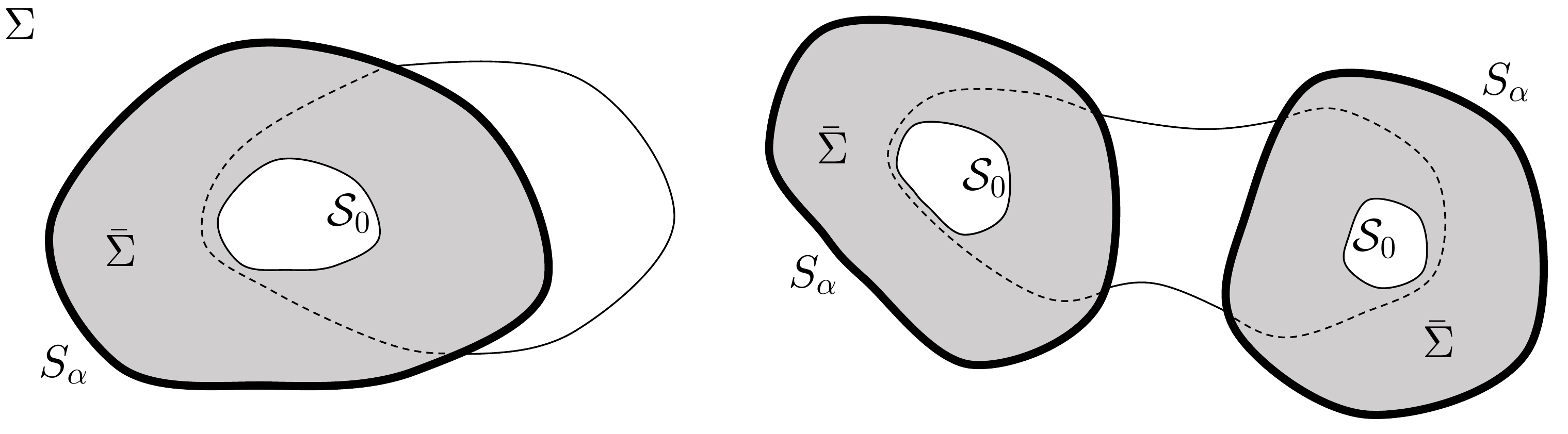}
  \end{center}
  \caption{Examples of the third case
    where ${\cal S}_0$ is not the outermost minimal surface: 
      The thick solid curves show the surface $S_\alpha$. 
      The outside of $S_\alpha$ is $\Sigma$ and the shaded region is $\bar\Sigma$. 
      The other boundary of $\bar\Sigma$ inside of $S_\alpha$
      is the minimal surface ${\cal S}_0$. 
      Minimal surfaces in $\Sigma$ touching with $S_\alpha$,
      which possibly exist, are shown by solid curves. They 
      might be further
      extended into $\bar\Sigma$ as described with dashed curves. 
      If there exists such a smooth closed minimal surface
      with a vanishing mean curvature,
      ${\cal S}_0$ is not the outermost minimal surface.}
   \label{FigMS}
\end{figure}

The last possibility
may occur in the following situation. 
In general, there 
exist minimal surfaces touching with the boundary of $\bar\Sigma$. 
If one of them can be extended to a compact minimal surface in $\Sigma$, 
we could have a minimal surface enclosing the minimal surface at $r=-r_0+2\log 2$ (see Fig.~\ref{FigMS}). 
If this occurs, we cannot apply Bray's theorem to the minimal surface at $r=-r_0+2\log 2$.

In our theorem, however, we assume that no minimal surface exists in $\Sigma$ whose area is less than $4\pi (2Gm)^2$ and which has a boundary on $S_\alpha$. 
This prevents the situation in Fig.~\ref{FigMS} for the following reason. 
If we have a minimal surface crossing $S_\alpha$ like the one shown in Fig.~\ref{FigMS}, 
its area has to be smaller than or equal to $4\pi (2Gm)^2$ by Bray's theorem. 
However, by assumption, the area of its part 
existing in the domain $\Sigma$ is equal to or larger than $4\pi (2Gm)^2$. 
Hence, the area of the connected minimal surface becomes larger than $4\pi (2Gm)^2$, 
which contradicts Bray's theorem. 
This means that such a surface cannot exist. 

Summarizing all the  discussions above together, 
the minimal surface at $r=-r_0+2\log 2$ is
guaranteed to be the outermost one 
in the $C^0$-class manifold constructed in Sect.~\ref{SoP}. 
In the next subsection, a sequence of smooth extensions converging to the $C^0$-class manifold is constructed.
The applicability of the results in this section to them is discussed at the end of Sect.~\ref{SE}.


\subsection{Smooth Extension}\label{SE}

To apply Bray's proof for the Riemannian Penrose inequality,
the manifold should be smooth~\cite{Bray}. 
However, the constructed manifold $\Sigma \cup \bar \Sigma$ in
Sect.~\ref{SoP} is $C^0$-class at the gluing surface, generally. 
Here, we show that there is a sequence of
smooth manifolds with a non-negative Ricci scalar ${}^{(3)}R$ 
which uniformly converges to $\Sigma \cup \bar \Sigma$. 
On each smooth manifold, we can apply 
the Riemannian Penrose inequality, and taking the limit, 
we achieve the inequality of Eq.~\eq{PE} on $\Sigma \cup \bar \Sigma$.

\begin{figure}[tb]
  \begin{center}
    \includegraphics[width=15.0cm,bb=0 0 1045 654]{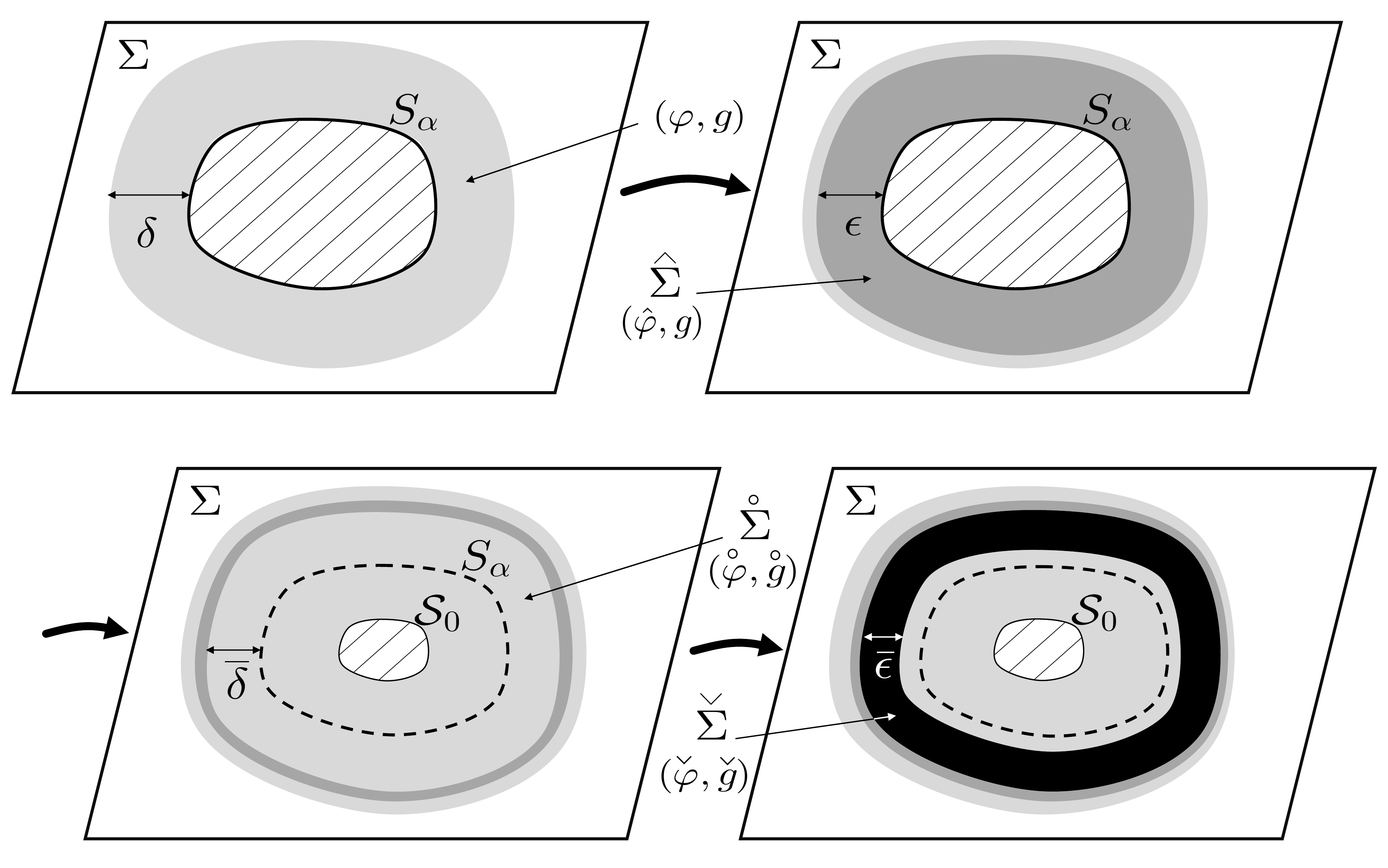}
  \end{center}
  \caption{Smooth deformation and extension of the manifold.
    There are three procedures. 
        The upper-left is the original manifold 
        $\Sigma$.
        The shaded region is the neighborhood of $S_\alpha$ ($\delta>r>0$)
        where the local inverse mean curvature flow can be taken. 
        In the first procedure,
        the lapse function is smoothly deformed in $\delta>\epsilon>r>0$,
        which is described in the upper-right, to make the
        Ricci scalar ${}^{(3)}R$ strictly positive.
        The obtained manifold is described by $\hat{\Sigma}$
        with the metric $(\hat{\varphi},g)$.
        In the second procedure, a manifold $\mathring{\Sigma}$
        with a metric $(\mathring{\varphi}, \mathring{g})$
        is attached at $r= \bd$ ($\epsilon>\bd> 0$)
        to obtain the lower-left manifold. 
        Since the method of gluing the two manifolds
        is the same as that in Sect.~\ref{SoP}
        except that the gluing position is slightly shifted, 
        the glued manifold is $C^0$-class at $r=\bd$. 
        In the third procedure, we smooth the glued manifold by introducing
        a manifold $\check{\Sigma}$ with the metric
        $(\check{\varphi}, \check{g})$ in the domain $\bd>r>\bd-\be>0$, 
        and a smooth manifold in the lower-right is obtained. 
      }
   \label{Figex}
\end{figure}

\subsubsection{Brief sketch}
First, without details, we give a brief sketch on
the smooth deformation and extension of the manifold (see Fig. \ref{Figex})
for the proof of Theorem~\ref{theorem}. 
Although Fig.~\ref{Figex} shows one component of $S_\alpha$,
the smooth deformation and extension should be done for all components of $S_\alpha$ simultaneously.
We take the foliation of the local inverse mean curvature flow with the
metric of Eq.~\eq{wohatm} in the neighborhood of $S_\alpha$ ($\delta>r>0$). 
The first procedure is to
deform the lapse function $\varphi$ to $\hat \varphi$ smoothly
in $\delta>\epsilon>r>0$ to make 
the three-dimensional Ricci scalar strictly positive.
The deformed manifold is denoted by $\hat{\Sigma}$ and  
its deformed metric is written as $(\hat\varphi,g)$.
The second procedure is to consider the $C^0$-extension of the manifold
in a similar way to that in Sect.~\ref{SoP}, 
but the position at which the two manifolds are glued is
slightly shifted: We choose the gluing position to be
$r=\bd$ ($\epsilon>\bd>0$), not $r=0$. 
The extended manifold is denoted by $\mathring{\Sigma}$ with the metric $(\mathring{\varphi},\mathring{g})$, and 
it uniformly converges to $\bar\Sigma$. 
As seen in Sect.~\ref{SoP}, $\mathring{\Sigma}$ has the non-negative
Ricci scalar everywhere. 
The third procedure is to  introduce a manifold $\check{\Sigma}$ in
the domain 
$\bd>r>\bd-\be>0$, where the metric $(\check{\varphi}, \check{g})$ 
is given by the mixture of
$(\hat\varphi,g)$ of $\hat \Sigma$ and $(\mathring{\varphi},\mathring{g})$ of $\mathring{\Sigma}$ so that $\check{\Sigma}$ 
is smoothly connected to $\hat{\Sigma}$
and $\mathring{\Sigma}$ 
at $r=\bd$ and $r=\bd-\be$, respectively. 
Then, we can show that $\check{\Sigma}$ with
the metric $(\check{\varphi}, \check{g})$ has
the positive Ricci scalar everywhere. 
Therefore, the obtained manifold,
which is the combination of the original manifold in $r\ge\epsilon$,
$\hat{\Sigma}$ with the metric $(\hat\varphi,g)$ in $\epsilon >r\ge \bd$,
$\check{\Sigma}$ with the metric $(\check{\varphi}, \check{g})$ in $\bd>r>\bd-\be$,
and 
$\mathring{\Sigma}$ with the metric $(\mathring{\varphi},\mathring{g})$ in $\bd-\be\ge r$,
is smooth, and has the positive Ricci scalar everywhere. 
The statements in the previous subsection are applied to the boundary of the deformed region, {\it i.e.} the surface $\cS_r$ at $r=\epsilon$ (denoted by $\cS_\epsilon$), not to $S_\alpha$ (at $r=0$).
The sequence of smooth manifolds
uniformly converges to $\Sigma \cup \bar \Sigma$
in the limit $\epsilon \to 0$.
Below, we provide a detailed description. 
The proof of Theorem~\ref{theorem2} can be achieved via a minor change to that of Theorem~\ref{theorem}.


\subsubsection{The first procedure}
We now begin with the detailed proof of Theorem~\ref{theorem}.
Since the original manifold $\Sigma$ is smooth,
we can take a smooth local inverse mean curvature flow~\cite{H&I,HP},
where $k$ is also a smooth function 
in the neighborhood of $S_\alpha$. 
Because of the positivity of $k$ on $S_\alpha$,
$k$ is positive in a sufficiently small neighborhood, 
$0\le r < \delta$. 
Similarly, since $r^a D_a k$ and  $k_{ab}$ are smooth as well,
there exists a positive constant $\beta$ 
satisfying 
\begin{eqnarray}
&&0<2r^a D_a k +  k^2 + k_{ab}^2 < \beta k^2 \label{defbeta}
\end{eqnarray}
everywhere in the neighborhood $0\le r < \delta$,
where we used our assumption, 
$r^a D_a k \ge \alpha k^2 > -(1/2) k^2$ on $S_\alpha$. 

On $S_\alpha$, the three-dimensional Ricci scalar $^{(3)}R$ is non-negative by assumption. 
The first procedure is to deform the metric so that $^{(3)}R$ becomes
strictly positive on and near $S_\alpha$. 
If it is already satisfied, this first procedure is unnecessary
and we can skip to the second procedure.
Otherwise, 
we introduce another metric in the region $0 \le r < \epsilon < \delta$ as 
\begin{eqnarray}
d\hat s ^2 &=\hspace{1.5mm}& u^2(r) \varphi^2 d r^2 + g_{ab} dx^a dx^b
 \nonumber \\
&=:& \hat  \varphi^2 d r^2 + g_{ab} dx^a dx^b, 
 \label{hatm} \\
u(r) &:=& 1 - \exp\left( -\frac{C \delta}{\epsilon -r }\right),
\end{eqnarray}
where $C$ is a constant.
At $r=\epsilon$, this metric is smoothly connected
with the original metric of Eq.~\eq{wohatm}.
The connected smooth manifold is denoted by $\hat{\Sigma}$. 
This metric uniformly converges to the metric
of Eq.~\eq{wohatm} in the limit $\epsilon \to 0$. 
The geometrical quantities with the metric of Eq.~\eq{hatm} on
an $r$-constant surface are written as 
\begin{eqnarray}
&&{}^{(2)}\hR = {}^{(2)} R , \qquad \hat \varphi^{-1} \hat{\cal D}^2 \hat \varphi =  \varphi^{-1} {\cal D}^2  \varphi , \nonumber \\
&&\hk_{ab} = u^{-1} k_{ab}, \qquad \hk = u^{-1} k, \qquad \hat{r}^a \hat{D}_a \hat{k} = - u^{-2} k^2 (\partial_{r} \log u) + u^{-2} r^aD_a k,
\end{eqnarray}
where we used $\varphi k =1$ in the last equation. 
Here, the variables with and without hat, such as $\hk$ and $k$,
indicate those of the metrics of Eqs.~(\ref{hatm}) and (\ref{wohatm}),
respectively. 
The three-dimensional Ricci scalar of the metric of Eq.~\eq{hatm}
can be expressed as
\begin{eqnarray}
{}^{(3)}\hR &=&  {}^{(2)}\hR - 2 \hat \varphi^{-1} \hat{\cal D}^2 \hat \varphi - 2 \hat{r}^a \hat{D}_a \hat{k} -  \hk^2 -  \hk_{ab}^2 
\nonumber \\ 
&=& {}^{(3)}R +\left(1-u^{-2}\right)  \left(  2 r^a D_a k +  k^2 + k_{ab}^2\right) + 2 u^{-2}k^2 \partial_{r} \log u.
\end{eqnarray}
The functions involving $u$ in the above equation are estimated as 
\begin{eqnarray}
0&>& 1-u^{-2}  
\ =\ 
u^{-2} \left[-2 \exp\left( -\frac{C\delta}{\epsilon - r}\right) + \exp\left( -2\frac{C\delta}{\epsilon - r}\right) \right]  \nonumber \\
&>& -2 u^{-2} \exp\left( -\frac{C\delta}{\epsilon - r}\right),
\end{eqnarray}
\begin{eqnarray}
\partial_{r} \log u &=& u^{-1}  \frac{C\delta}{(\epsilon - r)^2} \exp\left( -\frac{C\delta}{\epsilon - r}\right)
\ >\ 
\frac{C}{\delta} \exp\left( -\frac{C \delta}{\epsilon - r}\right).
\end{eqnarray}
These relations, together with Eq.~(\ref{defbeta}), imply 
\begin{eqnarray}
{}^{(3)}\hR \ >\ {}^{(3)}R + 2 \frac{k^2}{u^2} \left(\frac{C}{\delta} - \beta\right) \exp\left( -\frac{C\delta}{\epsilon - r}\right) .
\end{eqnarray}
Therefore, if we take $C> \beta \delta> 0$,  ${}^{(3)}\hR$ is strictly positive in the region $0 \le r < \epsilon$. 
On $S_\alpha$, $(\hat{r}^a \hat{D}_a \hat{k})/\hk^2$ is estimated as
\begin{eqnarray}
\left. \frac{\hat{r}^a \hat{D}_a \hat{k}}{\hk^2} \right|_{S_\alpha} &=&  \left. \frac{ (r^a D_a k)}{k^2} \right|_{S_\alpha} -  \left. \partial_r \log u  \right|_{S_\alpha}
\nonumber \\
&=& \left. \frac{ (r^a D_a k)}{k^2} \right|_{S_\alpha} - \dfrac{C\delta}{\epsilon^2} \frac{ \exp \left(- \dfrac{C\delta}{\epsilon} \right)}{1-\exp \left(- \dfrac{C\delta}{\epsilon} \right)} \nonumber \\
&\ge & \alpha + C_1 \epsilon =: \bar\alpha,
 \label{def:bar_alpha}
\end{eqnarray}
where $C_1$ is a (negative) constant. 

\subsubsection{The second procedure}
We now describe the second procedure.
We introduce a manifold $\mathring{\Sigma}$
with the following metric: 
\begin{eqnarray}
d \mathring{s}^2 &=&  \frac{ 1-2\exp \left( -\frac{r_{\bd}}{2} \right)}{1-2\exp \left( -\frac{r-\bd+r_\bd}{2} \right)}\exp \left( r -\bd \right) \hat\varphi_\bd^2 (x^a) dr^2 
+\exp \left( r -\bd \right) g_{\bd,ab} (x^a) dx^a dx^b 
\nonumber \\
&=:& \mathring{\varphi}^2 dr^2 + \mathring{g}_{ab} dx^a dx^b, 
\label{bdm}
\end{eqnarray}
where 
$\hat\varphi_\bd = \hat\varphi|_{r=\bd}$ and $g_{\bd,ab} (x^a) = g_{ab}|_{r=\bd}$.
Here, this metric has one parameter $\bd$, which
is required to satisfy $0< \bd < \epsilon$,
and $r_\bd$ is determined later.  
This manifold $\mathring{\Sigma}$
uniformly converges to $\bar\Sigma$ in the limit
$\epsilon \to 0$ (which leads to $\bd \to 0$).
We glue the two manifolds $\hat{\Sigma}$
and $\mathring{\Sigma}$ along the surface $r=\bd$ denoted by $\cS_\bd$.
Since $\hat{r}^a \hat{D}_a \hat{k}$ and $\hat k$ are smooth regular functions in $0<r\le \bd$, 
the relation between the values of
$\hat{r}^a \hat{D}_a \hat{k}/\hat{k}^2$
on $\cS_\bd$ and on $S_{\alpha}$ is estimated as 
\begin{eqnarray}
\left. \frac{\hat{r}^a \hat{D}_a \hat{k}}{\hat k^2 }\right|_{\cS_\bd} &\ge &
 \left. \frac{\hat{r}^a \hat{D}_a \hat{k}}{\hat k^2 }\right|_{S_\alpha} + \bar C_2 \epsilon \nonumber \\
&\ge & \bar\alpha + \bar C_2 \epsilon 
 =  \alpha + C_2 \epsilon =: \mathring{\alpha},
\end{eqnarray}
where $\bar C_2$ and $C_2(=C_1+\bar C_2)$ are constants,
and $\bar{\alpha}$ is determined in Eq.~\eqref{def:bar_alpha}.
Note that $\mathring{\alpha}>- 1/2$ can be satisfied by taking
a sufficiently small $\epsilon$, 
because $C_2$ is a constant independent of $\epsilon$.\footnote{
 This is required for $\mathring{\Sigma}$ to  converge uniformly to $\bar \Sigma$.}
Similarly to the case of the metric of Eq.~\eq{barm} in Sect.~\ref{SoP}, 
$\mathring{\varphi}$ and $r_\bd$ are taken
so that 
$\mathring{k} = \hk$ and 
$(\mathring{r}^a \mathring{D}_a \mathring{k})  = \mathring{\alpha} \mathring{k}^2$ are satisfied on $\cS_{\bd}$. 
Here, the geometric quantities with
circles (such as $\mathring{k}$) are those with respect to the metric
of Eq.~\eq{bdm}. 
At this point, the obtained manifold, $\mathring{\Sigma}$ glued to $\Sigma$ on $\cS_\bd$, is still $C^0$-class, generally.

\subsubsection{The third procedure}
The third procedure
is as follows. 
Let us deform the glued manifold in the domain $\bd - \be < r < \bd$ to make
it smooth, 
where $\be$ is any constant satisfying $0<\be<\bd$. 
We consider the difference of the metrics of Eqs.~\eq{hatm} and \eq{bdm}
around $\cS_\bd$. 
On $\cS_\bd$, the components of the two metrics have been set
to be equal to each other.
Since $\hat{k}$ and $\mathring{k}$ are equal on $\cS_{\bd}$, 
the first-order derivative of the metric component proportional
to $g_{ab}|_{r=\bd}$ is continuous as well.
By contrast, the first-order derivative
of the traceless components of the metric is not continuous. 
Therefore, since each metric is smooth,
the difference between them can be written as 
\begin{eqnarray}
 g_{ab} - \mathring{g}_{ab} =  (r-\bd)^2 T \mathring{g}_{ab} + (r-\bd)  h_{ab}, \qquad \hat \varphi- \mathring{\varphi}  = 2 (r-\bd) \mathring{\varphi} \Phi,
\end{eqnarray}
with smooth functions $T$, $h_{ab}$ and $\Phi$,
where $h_{ab}$ is traceless, and $\mathring{g}^{ab} h_{ab}=0$. 
In $\bd-\be< r < \bd$, we consider a manifold with the
following smoothed metric:
\begin{eqnarray}
d \check s^2 &=& \mathring{\varphi}^2 \left( 1 + \left[\frac{d}{dr} F^2(r) \right] \Phi\right)^2 dr^2 
+
\left(
\mathring{g}_{ab} \left[ 1 +  F^2(r)  T \right] + F(r) h_{ab}
\right) dx^a dx^b,
\nonumber \\
&=:& \check \varphi ^2 dr^2 + \check g_{ab} dx^a dx^b, 
\label{chem}
\end{eqnarray}
where 
\begin{eqnarray}
F(r) := (r-\bd)f(r-\bd)
\end{eqnarray}
with a function $f(r)$ defined on $-\be<r<0$ that satisfies
\begin{eqnarray}
&&\lim_{r \nearrow 0} f(r)= 1, \quad \lim_{r \searrow -\be} f(r) =0, \quad \lim_{r \nearrow 0}  f^{(n)} (r)= \lim_{r \searrow -\be} f^{(n)} (r)=0, 
\nonumber \\
&&
\left| \frac{d}{dr} (r f) \right| <1, \quad \frac{d^2}{dr^2} (r f)^2  <2. \label{fcond}
\end{eqnarray}
Here, $f^{(n)}$ is the $n$th order derivative of
$f$ for an arbitrary positive integer $n$.
Note that there exist functions satisfying the above requirements;
we show an example in Appendix~\ref{App}. 
Then, this manifold can be smoothly glued to $\Sigma$ with
the metric of Eq.~\eq{hatm} on $\cS_{\bd}$ and $\mathring{\Sigma}$ with the 
metric of Eq.~\eq{bdm} at $r= \bd-\be$. 
Moreover, the obtained manifold uniformly converges
to $\Sigma \cup \bar\Sigma$ with the metrics of Eqs.~\eq{wohatm} and \eq{barm} 
in Sect.~\ref{SoP} in the limit $\epsilon \to 0$ (which gives $\bd \to 0$ and $\be \to 0$). 
The metric of Eq.~\eq{chem} and its derivatives
with respect to $r$ are estimated
in terms of those of the metric of Eq.~\eq{hatm}, as 
\begin{eqnarray}
&&\check \varphi = \mathring{\varphi}|_{r=\bd} + \cO(\bd-r), \qquad
\check g_{ab} = \mathring{g}_{ab}|_{r=\bd} + \cO(\bd-r),  
\nonumber \\
&&
\check g^{ab} = \mathring{g}^{ab}|_{r=\bd} - F h^{ab} + \cO\left((\bd-r)^2 \right), 
\nonumber \\
&&
\partial_ r \log \check \varphi = (\partial_r \log \mathring{\varphi})|_{r=\bd} + (F^2)'' \Phi+ \cO(\bd-r), 
\\
&&
\partial_ r \check g_{ab} = (\partial_r \mathring{g}_{ab})|_{r=\bd} + (F)' h_{ab}+ \cO(\bd-r), 
\nonumber \\
&&
\partial_ r ^2 \check g_{ab} = \left. \left( \partial_r^2 \mathring{g}_{ab}\right)\right|_{r=\bd} 
+ \mathring{g}_{ab} \left(F^2 \right)'' T + F'' h_{ab} + 2 F' \partial_r h_{ab}+ \cO(\bd-r), \nonumber
\end{eqnarray}
where the prime means the derivative with respect to $r$, {\it i.e.}, $F' = dF/dr$, 
and $h^{ab}:= \mathring{g}^{ac} h_{cd} \mathring{g}^{db}$.
Note that the order of $F$, $F^2$ and their derivatives are
 $F=\cO(\be)$, $F^\prime = \cO(1)$, $F^{\prime\prime} = \cO(1/\be)$,
$F^2=\cO(\be^2)$, $(F^2)^\prime = \cO(\be)$, and $(F^2)^{\prime\prime} = \cO(1)$.
Then, the geometric quantities are expressed as 
\begin{eqnarray}
&&\check k_{ab} = \mathring{k}_{ab}|_{r=\bd} + \frac{F' }{2 \mathring{\varphi}|_{r=\bd}} h_{ab} + \cO(\bd-r), \qquad \check k =  \mathring{k}|_{r=\bd}+ \cO(\bd-r), \nonumber \\
&&
\check r^a \check D_a \check k = (\mathring{r}^a \mathring{D}_a \mathring{k}) |_{r=\bd} + (F^2)'' \left.\left(\frac{4T-4 \mathring{\varphi} \Phi  \mathring{k}  - h_{ab}^2}{4 \mathring{\varphi}^2 }
\right)\right|_{r=\bd} + \cO(\bd-r).
\\
&&
 {}^{(2)} \check R = {}^{(2)} \mathring{R}|_{r=\bd} + \cO(\bd-r), \qquad \check \varphi^{-1} \check {\cal D}^2 \check \varphi =  \left( \mathring{\varphi}^{-1} \mathring{\cal D}^2 \mathring{\varphi} \right) \Big|_{r=\bd} + \cO(\bd-r) , \nonumber
\end{eqnarray}
where the geometric functions with check marks (such as $\check{k}$)
are those with respect to the metric of Eq.~\eq{chem}.
Note that we used the traceless condition of $h_{ab}$,
{\it i.e.} $\mathring{g}^{ab} h_{ab}=0$, to simplify the expressions.
The three-dimensional Ricci scalar for the metric of Eq.~\eq{chem}
is reduced to 
\begin{eqnarray}
  &&{}^{(3)} \check R = {}^{(3)} \mathring{R}|_{r=\bd} - F'^2  \left. \left( \frac1{2 \mathring{\varphi}} h_{ab} \right)^2 \right|_{r=\bd} - \left(F^2\right)''
  \left.\left(\frac{4T-4 \mathring{\varphi} \Phi  \mathring{k}  - h_{ab}^2}{2 \mathring{\varphi}^2 }
\right)\right|_{r=\bd}
\nonumber \\
&&
\hspace{110mm}
+ \cO(\bd-r),
\end{eqnarray}
where we used $\mathring{k}^{ab}h_{ab}=0$
since $r=\mathrm{constant}$
surfaces are umbilical
for the metric of Eq.~\eqref{bdm}.
Here, we evaluate the value of
the three-dimensional Ricci scalar ${}^{(3)}\hat{R}$ for the metric
of Eq.~\eq{hatm} ({\it i.e.} on $\hat{\Sigma}$)
on $r=\bd$. Since ${}^{(3)}\check{R}$ and ${}^{(3)}\hat{R}$
have the same value on $r=\bd$, 
we have
\begin{eqnarray}
{}^{(3)} \hat R|_{r=\bd} = {}^{(3)} \mathring{R}|_{r=\bd} -   \left. \left( \frac1{2 \mathring{\varphi}} h_{ab} \right)^2 \right|_{r=\bd} 
-
\left.\left(\frac{4T-4 \mathring{\varphi} \Phi  \mathring{k}  - h_{ab}^2}{\mathring{\varphi}^2 }
\right)\right|_{r=\bd},
\end{eqnarray}
using $F^\prime = 1$ and $(F^2)^{\prime\prime}=2$ at $r=\bd$. 
In the same way, we have 
\begin{eqnarray}
\left. \left(\hat{r}^a \hat{D}_a \hat{k}\right) \right|_{r=\bd} =\left.\left(\mathring{r}^a \mathring{D}_a \mathring{k}\right)\right|_{r=\bd} +
\left.\left(\frac{4T-4 \mathring{\varphi} \Phi  \mathring{k}  - h_{ab}^2}{2 \mathring{\varphi}^2 }
\right)\right|_{r=\bd}.
\end{eqnarray}
Since
\begin{eqnarray}
\left. \frac{\hat{r}^a \hat{D}_a \hat{k}}{\hat k^2} \right|_{r=\bd} \ge \mathring{\alpha} = 
\left. \frac{\mathring{r}^a \mathring{D}_a \mathring{k}}{\mathring{k^2} }\right|_{r=\bd} 
\qquad \mbox {and} \qquad 
 \hat k=\mathring{k}
\end{eqnarray}
are satisfied from the construction of the metric of Eq.~\eq{bdm}, we have 
\begin{eqnarray}
\left( 
4T-4 \mathring{\varphi} \Phi \mathring{k}  - h_{ab}^2
\right) \Big|_{r=\bd} \ge 0.
\end{eqnarray}
Using the above and the inequalities of Eq.~\eq{fcond}, we find
\begin{eqnarray}
{}^{(3)} \check R &=&
{}^{(3)} \hat R|_{r=\bd}+ \left(1 - F'^2 \right)
\left. \left( \frac1{2 \mathring{\varphi}} h_{ab} \right)^2 \right|_{r=\bd} 
\nonumber \\
&&\hspace{10mm}
+ \left[2-\left( F^2\right)'' \right]
\left.\left(\frac{4T-4 \mathring{\varphi} \Phi  \mathring{k}  - h_{ab}^2}{2 \mathring{\varphi}^2 }
\right)\right|_{r=\bd}
+ \cO(\bd-r) 
\nonumber \\
&\ge& {}^{(3)} \hat R|_{r=\bd} + \cO(\be). 
\end{eqnarray}
Since ${}^{(3)} \hat R|_{r=\bd}$ is strictly positive,
${}^{(3)} \check R$ can be made positive everywhere
by setting $\be$ to be a sufficiently small value.
In other words, 
there is a critical value $\be_c$ such that
${}^{(3)} \check R$ becomes non-negative for $\be\le\be_c$. 
As a result, there exists a sequence of smooth manifolds with
a non-negative Ricci scalar, which uniformly converges to 
$\Sigma \cup \bar\Sigma$ with the metrics of Eqs.~\eq{wohatm} and \eq{barm}. 

\subsubsection{Reconsideration of Sect.~\ref{OMMS}}
We should confirm that
the discussion in Sect.~\ref{OMMS} is applicable in the sequence of the smooth manifolds constructed above. 
If the boundary of the deformed region is $\cS_\epsilon$ (see Fig.~\ref{fig:T1}), 
the discussion in Sect.~\ref{OMMS} should be applied for $\cS_\epsilon$, instead of $S_\alpha$. 
For $r<\epsilon$ in any constructed smooth manifold, the mean curvature
$k$ on each $r$-constant surface is positive. 
By taking  $\epsilon<\varepsilon$, where $\varepsilon$ is introduced in Theorem~\ref{theorem},  
the surface $\cS_\epsilon$ at $r=\epsilon$ does not have minimal surfaces satisfying conditions {\it (i)} and {\it (ii)} by the assumption of Theorem~\ref{theorem}. 
Therefore,
the result in Sect.~\ref{OMMS} holds for the constructed smooth manifold. 
Then, by taking the limit $\epsilon \to 0$,  the proof of Theorem~\ref{theorem} is completed.

\subsubsection{Proof of Theorem.~\ref{theorem2}}
\begin{figure}[tb]
 \begin{minipage}{0.49\hsize}
  \begin{center}
   \includegraphics[width=70mm]{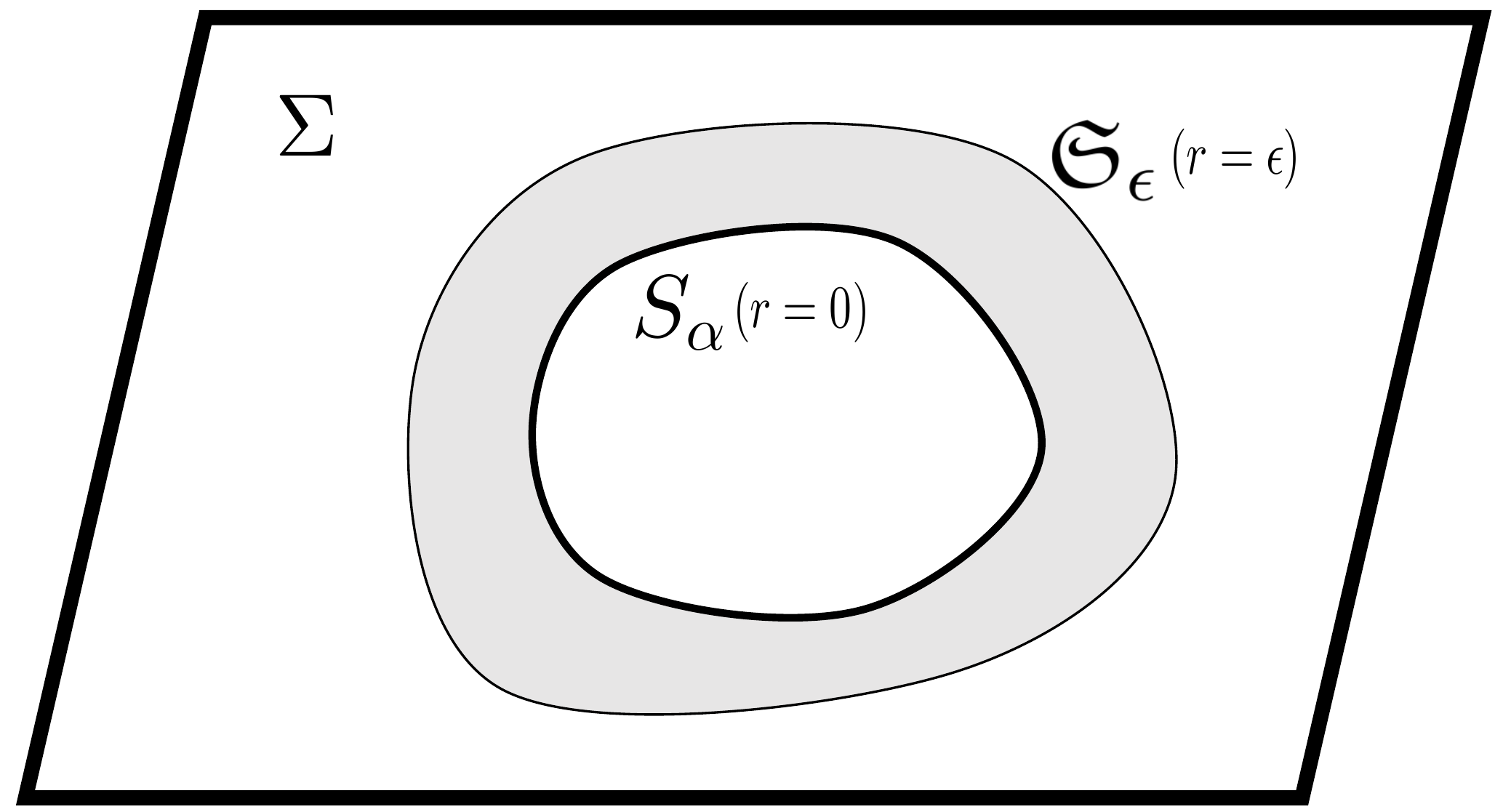}
  \end{center}
  \caption{Deformed region in  Theorem~\ref{theorem}: The deformed region is shown as the gray one. The thick and thin curves correspond to $S_\alpha$ ($r=0$) and  $\cS_\epsilon$ ($r=\epsilon$), respectively.
  The local mean curvature flow is taken in exterior of $S_\alpha$. The smooth deformation is done for $0<r<\epsilon$, and the outer boundary of the deformed region is located at $r=\epsilon$, that is $\cS_\epsilon$. }
  \label{fig:T1}
 \end{minipage}
 \hspace{0.02\hsize}
 \begin{minipage}{0.49\hsize}
  \begin{center}
   \includegraphics[width=70mm]{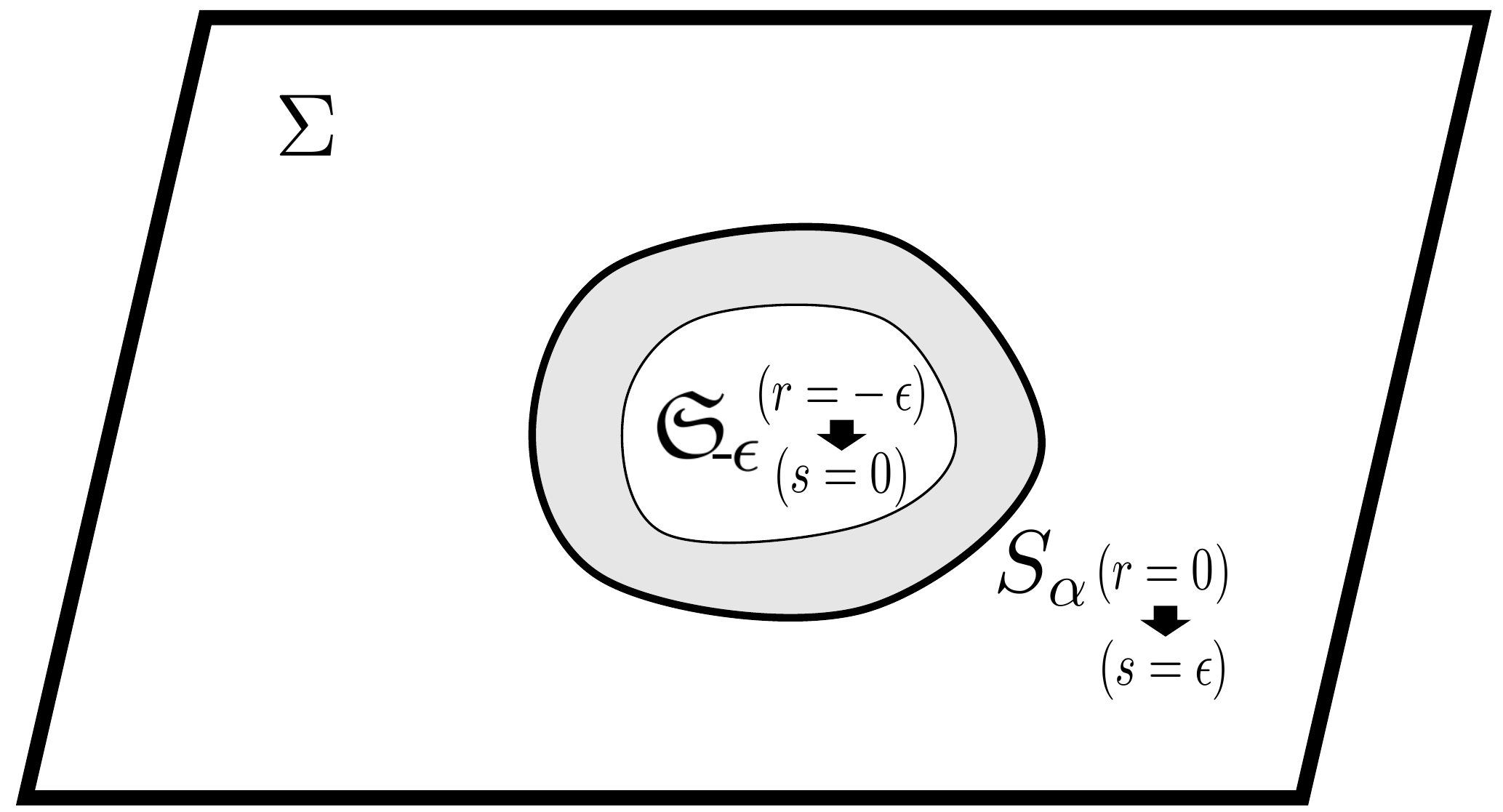}
  \end{center}
  \caption{Deformed region in  Theorem~\ref{theorem2}: The gray-colored part shows the deformed region. The thick and thin curves correspond to $S_\alpha$ ($r=0$) and  $\cS_{-\epsilon}$ ($r=-\epsilon$), respectively. 
  By assumption, the local mean curvature flow can be taken in the interior of $S_\alpha$ ($-\epsilon<r<0$). 
  The smooth deformation can be done there and the boundary of deformed region can be $S_\alpha$.}
  \label{fig:T2}
 \end{minipage}
\end{figure}

With respect to Theorem~\ref{theorem2}, no assumptions for each leaf of the foliation  exist. 
Instead, a  smooth inward extension from $S_\alpha$ is assumed to exist, and 
the local mean curvature flow is also assumed to be taken in the interior of $S_\alpha$. 
In the proof of Theorem~\ref{theorem}, the deformation occurs in the range $0<r<\epsilon$, 
where the local inverse mean curvature flow is taken, and 
the outer boundary of the smooth deformation is  located at $r=\epsilon$, {\it i.e.}, $\cS_\epsilon$.
In the setting of Theorem~\ref{theorem2}, on the other hand, 
since the existence of the local mean curvature flow in the interior of $S_\alpha$, {\it i.e.}, $r<0$, is assumed,  
the deformation based on the local mean curvature flow can be done in the range $-\epsilon<r<0$
(see Fig.~\ref{fig:T2} and compare between Figs.~\ref{fig:T1} and \ref{fig:T2}). 
Then, its outer boundary is located exactly at $r=0$, {\it i.e.}, $S_\alpha$. 
Such a deformation can be achieved 
in the same way as that of Theorem~\ref{theorem}, 
but
the coordinate $r$ in  the proof of Theorem~\ref{theorem} is replaced by
a new coordinate $s$ defined as $r= s - \epsilon$.
The analysis is carried out by replacing $r$ in the proof of Theorem~\ref{theorem} by $s$, 
thus the outer boundary of the deformed region is located at $s = \epsilon$, which is $S_\alpha$. 
The result in Sect.~\ref{OMMS} is applicable without change, 
and taking the limit $\epsilon \to 0$ gives Theorem~\ref{theorem2}.

\section{Summary} \label{secSum}

In this paper, we have introduced a new concept,
the attractive gravity probe surface (AGPS), 
to characterize the strengths of gravitational field.
An AGPS is a two-dimensional closed surface that
satisfies $k>0$ and $r^a D_a k/ k^2 \ge \alpha$ 
for the local inverse mean curvature flow,
where $\alpha$ is a constant satisfying 
$\alpha>-1/2$. 
This concept can be applied to surfaces
in regions with various strength of the gravitational field,
including weak-gravity regions: 
The parameter $\alpha$ controls the strength of the gravitational field. 
For instance, for $r$-constant surfaces in
a $t$-constant slice of a Schwarzschild spacetime, 
$r^a D_a k/ k^2$ is a monotonically decreasing continuous function of $r$.
It converges to $\infty$ in the limit where the surface approaches
the horizon,
while it converges to $-1/2$ at spatial infinity. 
Hence, $\alpha$
is interpreted as an indicator for the minimum strength of the 
gravitational field on the surface, 
because it gives the lower bound of $r^a D_a k/ k^2$ on that surface.

Then, we have generalized the Riemannian Penrose inequality
to the areal inequality for AGPSs; 
for asymptotically flat or AdS spaces,
the area of an AGPS surface 
has an upper bound, $A_\alpha \le 4\pi [ ( 3+4\alpha)/(1+2\alpha) ]^2(Gm)^2$
(the exact statement is presented in Sect.~\ref{IMCFGI}
and as Theorem~\ref{theorem}
in Sect.~\ref{MT}). 
In the limit $\alpha \to \infty$ and in the case $\alpha =0$, 
our inequality is reduced to the Riemannian Penrose inequality and
its analog for the LTS, respectively. 
The proof for asymptotically flat spaces is carried out
by applying Bray's theorem derived using the conformal flow.
This means that our inequality is applicable
to the case of an AGPS with multiple components. 
For asymptotically AdS spaces, our argument relied on
the Geroch monotonicity in the inverse mean curvature flow. 
The proof for the areal inequality for AGPSs based on the conformal flow,
which requires the generalization of Bray's method
to AdS spaces, 
will be
an important remaining problem.  

It would be interesting to extend our inequality to that in
Einstein-Maxwell systems, 
as in the case of the Riemannian Penrose inequality~\cite{Jang:1979zz,Weinstein:2004uu,Khuri:2013ana,Khuri:2014wqa} and
its analog for the LTS~\cite{Lee:2020pre}. 
Moreover, the extension to the higher-dimensional cases
is expected to be possible for asymptotically flat spaces. 
We leave these topics for future works. 

It is also important to consider the relation to the behavior
of photons and the possible connection to observations.  
For LTSs, the relations to dynamically transversely trapping surfaces,
which is a generalization of the concept of the photon surface
for generic spacetimes based on the photon orbits, 
have been
discussed~\cite{Lee:2020pre,Yoshino:2017gqv,Yoshino:2019dty,Yoshino:2019mqw}. 
It would be interesting to explore the relation
between the AGPSs and the photon behavior.

\section*{Acknowledgement}

K.~I. and T.~S.  are supported by JSPS Grants-in-Aid for Scientific Research (A)(No.\,17H01091).
K.~I. is also supported by JSPS Grants-in-Aid for Scientific Research (B) (20H01902).
H.~Y. is supported by the Grant-in-Aid for
Scientific Research (C) (No. JP18K03654) from Japan Society for
the Promotion of Science (JSPS).
The work of H.Y. is partly supported by
Osaka City University Advanced Mathematical Institute
(MEXT Joint Usage/Research Center on Mathematics and Theoretical Physics 
JPMXP0619217849).

\appendix

\section{An example of the matching function $f$}\label{App}

\begin{figure}[tb]
  \begin{center}
    \includegraphics[width=10.0cm,bb= 0 0 415 297]{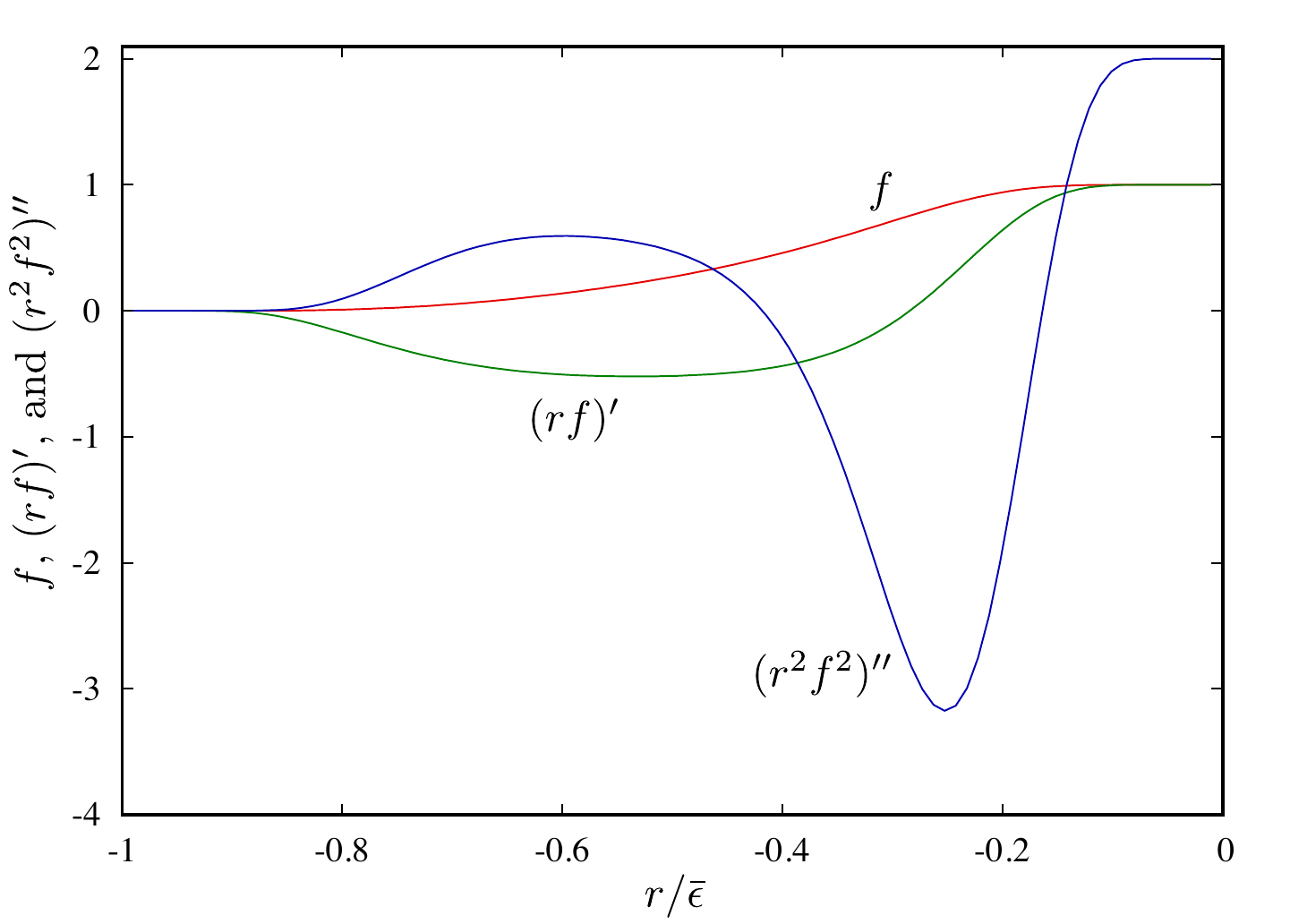}
  \end{center}
  \caption{ The behavior of the functions $f$, $(rf)^\prime$, and
    $(r^2f^2)^{\prime\prime}$ for the
    matching function given by Eq.~\eqref{example_matching_function}.
  }
   \label{Fig_smoothing_functions}
\end{figure}

An example of the matching function is
\begin{eqnarray}
  f (r) = \left( 1+ \exp x(r) \right)^{-1}, \qquad x(r):= \frac{\be}{r} +  \frac{\be}{r+\be} +1 ,
  \label{example_matching_function}
\end{eqnarray}
defined on $ -\be < r < 0$.
The behavior of $f$, $(rf)^\prime$ and $[(rf)^2]^{\prime\prime}$
is shown in Fig.~\ref{Fig_smoothing_functions},
where the prime means the derivative
with respect to $r$ (not to $x$),
{\it i.e.}, $f' = \frac{d}{dr} f \bigl( x(r)\bigr)$.
From this figure, one can expect
that the requirements of Eq.~\eqref{fcond} would be satisfied.
Here, we provide a solid proof for this fact.

\subsection{Evaluation of the first three equations of conditions of Eq.~\eq{fcond} }

It is trivial that the first two equations of Eq.~\eq{fcond}
are satisfied because of
\begin{eqnarray}
\lim_{r \nearrow 0} e^x= 0, \quad \lim_{r \searrow -\be} e^x =\infty.
\end{eqnarray}
Since the function $f$ converges to
$1$ and $0$ very rapidly in the limit of $r \nearrow 0$ and
$r \searrow -\be$, respectively,
the third equation of Eq.~\eq{fcond} is expected to be satisfied. 
Let us show it exactly.

The $n$th-order derivatives of $f(x(r))$ with respect to $r$
are written by the finite sum of terms, each of which has the form 
\begin{eqnarray}
c_l \frac{d^m f}{dx^m} \prod_{k=1}^{m} \frac{d^{a_k} x}{dr^{a_k}}, \label{df}
\end{eqnarray}
where $m$ is a positive integer,  $c_l$ is a constant
and the $a_k$ terms are positive integers satisfying 
\begin{eqnarray}
n= \sum_{k=1}^{m} a_k. 
\end{eqnarray}
Different $a_k$ terms may take the same integer. 
The third equation of Eq.~\eq{fcond} holds if each term
of Eq.~\eq{df} converges to zero in the limits
$r \nearrow 0$ and $r \searrow -\be$.

The $a_k$th-order derivatives of $x(r)$  become
\begin{eqnarray}
\dfrac{d^{a_k} x}{dr^{a_k}} = (-1)^{a_k} {a_k}! \left( \dfrac{1}{r^{{a_k}+1}} +\dfrac{1}{(r+\be)^{{a_k}+1}} \right) \be .
\end{eqnarray}
Around $r=0$, the absolute value of it is bounded from above as 
\begin{eqnarray}
\left| \dfrac{d^{a_k} x}{dr^{a_k}}\right| < c_{a_k} \left| \dfrac{1}{r^{{a_k}+1}}  \right| ,
\end{eqnarray}
where $c_{a_k}$ is a positive constant. 
Thus, the absolute value of
the product of the derivatives of $x$ in Eq.~\eq{df} is
bounded from above as 
\begin{eqnarray}
\prod_{k=1}^{m}\left|  \frac{d^{a_k} x}{dr^{a_k}} \right| < c_a \left| \dfrac{1}{ r^{a+m}}  \right|, 
\qquad \mbox{where} \qquad
c_a:= \prod_{k=1}^{m} c_k, \qquad a :=\sum_{k=1}^{m} a_k.
\end{eqnarray}

We now estimate the $m$th-order derivative of $f(x(r))$ with respect to $x$. 
The first-order derivative is calculated as
\begin{eqnarray}
\dfrac{df}{dx} = f(f-1) .\label{df1}
\end{eqnarray}
Defining a functional space $\cal F$ as 
\begin{eqnarray}
{\cal F}:=\left\{ f (f-1) p(f) \left| \mbox{$p(f)$ is a polynomial of $f$} \right. \right\},
\end{eqnarray}
we show 
\begin{eqnarray}
\frac{d^m f}{dx^m} \in {\cal F}   \label{cF}
\end{eqnarray}
for $m\ge 1$ by the method of induction. 
For $m=1$, it is true because of Eq.~\eq{df1}. 
Suppose Eq.~\eq{cF} is satisfied for $m=q$.
The assumption indicates that $d^q f / dx^q$ is written by $f (f-1) p_q(f)$ with a polynomial $p_q (f)$, 
and its derivative becomes
\begin{eqnarray}
\frac{d^{q+1} f}{dx^{q+1}} = f (f-1) \left[ (2f-1) p_q(f)+  \frac{d p_q (f)}{df} \right] \in {\cal F}.  
\end{eqnarray}
Therefore, Eq.~\eq{cF} is proved for arbitrary $m\ge 1$. 

Now, we know that each $m$th-order derivative is written as
\begin{eqnarray}
\frac{d^m f}{dx^m} = f (f-1) p_m(f), 
\end{eqnarray}
with a polynomial $p_m(f)$. 
The function $f$ converges to unity in the limit $r \nearrow 0$. 
Therefore, $|f p_m(f)|$ is bounded from above by a positive constant $\tilde c_m$ around $r=0$. 
On the other hand, $|f-1|$ is estimated around $r=0$ as 
\begin{eqnarray}
|f-1| = \frac{\exp x}{1+\exp x} < \bar c_m \exp x = \bar c_m \exp \left( \frac{\be}{r} +\frac{\be}{r+\be} +1 \right) < \bar c_m \exp \left( \frac{\be}{r} \right) ,
\end{eqnarray}
with a positive constant $\bar c_m$. 
Therefore, we have 
\begin{eqnarray}
\left |\frac{d^m f}{dx^m} \right| < \tilde c_m \bar c_m  \exp \left( \frac{\be}{r} \right) =: c_m \exp \left( \frac{\be}{r} \right),
\end{eqnarray}
around $r=0$. 
As a result, Eq.~\eq{df} is estimated around $r=0$ as 
\begin{eqnarray}
\left| c_l \frac{d^m f}{dx^m} \prod_{k=1}^{l} \frac{d^{a_k} x}{dr^{a_k}}\right| 
<  \left| c_l \right|  c_a c_m  \left| \dfrac{1}{ r^{a+l}}  \right| \exp \left( \frac{\be}{r} \right)
\xrightarrow[r \nearrow 0]{} 0 .
\end{eqnarray}
This leads to 
\begin{eqnarray}
 \lim_{r \nearrow 0}  f^{(n)} (r)= 0. 
\end{eqnarray}
In a similar way, we can show the vanishing of $f^{(n)} (r)$
for the limit $r \searrow -\be$.

\subsection{Estimate of $(r f)'$}
We now show that the condition $| (rf)' | <1$
is satisfied.
First, we show the upper bound of $(r f)'$ as follows:
\begin{eqnarray}
(rf)' &=&\dfrac{1}{1+e^x} \left[ 1+\dfrac{r \be e^x}{1+e^x} \left( \dfrac{1}{r^2} +\dfrac{1}{(r+\be)^2} \right) \right] 
< \dfrac{1}{1+e^x} 
\ < \ 1.
\end{eqnarray}
Note that $f(r)$ is defined on $-\be < r<0$,  {\it i.e.}, $r$ is negative.

Next, we estimate the lower bound of $(r f)'$ in the cases
of $-{\be}/{2} \le r <0$ and $-\be < r \le -{\be}/{2}$, separately.
It is useful to express $(r f)'$ as follows:
\begin{eqnarray}
(r f) ' &=& \frac{1}{ 1+ e^x } \left\{ 1 +
\frac{e^x}{ 1+ e^x } \left[ x - \left( \frac{\be}{\be+r} \right)^2 -1 \right] \right\} \label{rf1} \\ 
 &=& \frac{1}{ 1+ e^x } \left\{ 1 +
\frac{e^x}{ 1+ e^x } \left[ x - \left( x- \frac{\be}{r} -1\right)^2 -1 \right] \right\}. 
\label{rf2}
\end{eqnarray}

\subsubsection{The case $-{\be}/{2} \le r <0$}

In this case, we have the inequalities 
\begin{eqnarray}
x \le 1, \qquad e^x \le e , \qquad 1<\frac{\be}{\be+r} \le 2.
\end{eqnarray}
%
From Eq.~\eqref{rf1}, $(rf)'$ is bounded as
\begin{eqnarray}
 (r f)'  > \frac{1}{ 1+ e^x } \left[ 1 +  \frac{e^x}{ 1+ e^x }  (x-5) \right] .
\label{x-5}
\end{eqnarray}
The derivative of  the right-hand side of the above is 
\begin{eqnarray}
\dfrac{d}{d x}\left(\frac{1}{ 1+ e^x } \left[ 1 +  \frac{e^x}{ 1+ e^x }  (x-5) \right]  \right)
= \frac{e^x( 1-e^x )}{( 1+ e^x)^3 }   (x-5) .
\end{eqnarray}
This is negative for $x <0$ and positive for $0<x\le1$. 
Thus, the minimum of the right-hand side of Eq.~\eq{x-5}
occurs at $x=0$, and hence, we have 
\begin{eqnarray}
(r f)' \geq \frac{1}{ 1+ 1 } \left[ 1 +  \frac{1}{ 1+ 1 }  (-5) \right] = -\frac{3}{4} >-1 .
\end{eqnarray}

\subsubsection{The case $-\be < r \le -{\be}/{2}$}
In this case, we have
\begin{eqnarray}
x \ge 1, \qquad e^x \ge e , \qquad 1< - \frac{\be}{r} \le 2. 
\end{eqnarray}
From Eq.~\eq{rf2}, the lower bound is evaluated as
\begin{eqnarray}
 (r f)' 
&\ge&
\frac{1}{ 1+ e^x } \left\{ 1 +\frac{e^x}{ 1+ e^x } \left[ x - \left( x + 1 \right)^2 -1\right] \right\} \nonumber \\
&=& \frac{1}{ (1+ e^x)^2 } \left[ 1- e^x \left( x^2 + x +1 \right) \right]
\nonumber \\
&>& - \frac{1}{ (1+ e^{-x})^2 } e^{-x} \left( x^2 + x  +1 \right) .
\label{ddrf}
\end{eqnarray}
Since 
\begin{eqnarray}
&&\frac{d}{dx} \left[e^{-x} \left( x^2 + x  +1 \right)\right] 
=- e^x x(x-1) <0
\end{eqnarray}
holds, $e^{-x} \left( x^2 + x  +1 \right)$ is a decreasing function for $x>1$. 
Meanwhile, ${1}/{ (1+ e^{-x})^2 }$ is an increasing function. 
Then, $ (r f)'$ is found to be bounded from below as 
\begin{eqnarray}
 (r f)'  
>- \left[\frac{1}{ (1+ e^{-x})^2 }\right] \Big|_{x=2}  \left[e^{-x} \left( x^2 + x  +1 \right)\right] |_{x=1} >-1
\end{eqnarray}
for $1<x<2$, and 
\begin{eqnarray}
 (r f)'  
>- \left[\frac{1}{ (1+ e^{-x})^2 }\right] \Big|_{x\to\infty}  \left[e^{-x} \left( x^2 + x  +1 \right)\right] |_{x=2} >-1
\end{eqnarray}
for $2<x$.

\subsection{Estimate of $\left[ (r f)^2\right] ''$}
The second-order derivative of $ (r f)^2$ can be written as
\begin{eqnarray}
&& \left[ (r f)^2\right] ''=
\frac2{(1+e^x )^4} \left\{ (1+e^x )^2 + 2 e^x(1+e^x ) \left[ x-1- \left(\frac{\be}{\be+r}\right)^3\right]
\right. \nonumber \\ 
&& \hspace{55mm} \left.
+ e^x \left( 2 e^x -1 \right)  \left[ x-1- \left(\frac{\be}{\be+r}\right)^2\right]^2
\right\} .
\label{ddrrff}
\end{eqnarray}
We will show that $ \left[ (r f)^2\right] ''<2$ is satisfied,
in the three cases, $ e^x \le 1/2 $ ({\it i.e.} $x\leq -\log 2$), 
$1/2 < e^x \le e$ ({\it i.e.} $- 7/10 < -\log 2< x \leq 1$),
and $e < e^x $ ({\it i.e.} $1<x$), separately.

\subsubsection{The case $ e^x \le 1/2 $~~({\it i.e.} $x\leq -\log 2$)}

Since $x$ and $(2 e^x -1)$ are negative, Eq.~\eq{ddrrff} gives
\begin{eqnarray}
&& \left[ (r f)^2 \right] '' \le
\frac2{(1+e^x )^4}  (1+e^x )^2 <1.
\end{eqnarray}

\subsubsection{The case $1/2 < e^x \le e$~~({\it i.e.} $- 7/10 < -\log 2< x \leq 1$)}

We can express ${\be}/{r}$  in terms of ${\be}/{(r+ \be)}$ as
\begin{eqnarray}
\frac{\be}{r} = \frac{\be}{r+ \be} \left( 1-\frac{\be}{r+ \be}\right)^{-1},
\end{eqnarray}
and then, we can see a relation between ${\be}/{(r+ \be)}$ and $x$, 
\begin{eqnarray}
\left( \frac{\be}{r+ \be}\right)^2 = (x+1) \frac{\be}{r+ \be} -(x-1). \label{e/re}
\end{eqnarray}
The terms including ${\be}/{(r+ \be)}$ in the right-hand side
of Eq.~\eq{ddrrff} are rewritten as
\begin{eqnarray}
&&-\left( \frac{\be}{r+ \be}\right)^3 = - \left( x^2 +x +2 \right)  \frac{\be}{r+ \be} + x^2-1,
\\
&&\left[\left( \frac{\be}{r+ \be}\right)^2 -(x-1) \right]^2 = 
\left( x^3 - x^2 +3x +5 \right)  \frac{\be}{r+ \be} -x^3 +3x^2 -7x+5. 
\end{eqnarray}
Solving Eq.~\eq{e/re} for ${\be}/{(r+ \be)}$, it is estimated as
\begin{eqnarray}
\frac{7}{5}< \frac{\be}{r+ \be} = \frac{(x+1) + \sqrt{(x-1)^2+4}}{2} \leq 2 , \label{ineqre}
\end{eqnarray}
where we used $1<{\be}/{(r+ \be)}$ to choose the sign in the front of
the square root
and $-7/10< x\leq 1$. As a consequence, we find
\begin{eqnarray}
\left[ (r f)^2\right] ''&=&
\frac2{(1+e^x )^4} \biggl\{
1 + \left( x^3-x^2+9x -7 \right) e^x + \left( -2x^3+8x^2-12x+7 \right)e^{2x} 
 \nonumber \\ 
&&\hspace{10mm}
+ \left[  - \left( x^3+x^2+5x+9 \right) + \left(2x^3-4x^2+4x +6 \right)   e^x \right] e^x \frac{\be}{r+ \be}
\biggr\}
\nonumber \\
&=&
\frac2{(1+e^x )^4} \biggl\{
1 + \left( x^3-x^2+9x -7 \right)e^x + \left( -2x^3+8x^2-12x+7 \right)e^{2x} 
\nonumber \\ 
&&\hspace{20mm} 
+ \left[  - \left( 3x^2+3x+6 \right) + \left(x^3-2x^2+2x +3 \right)   (2e^x -1) \right] e^x \frac{\be}{r+ \be}
\biggr\}
\nonumber \\
&<&
\frac2{(1+e^x )^4} \biggl\{
1 + \left( x^3-x^2+9x -7 \right) e^x + \left( -2x^3+8x^2-12x+7 \right)e^{2x} 
\nonumber \\ 
&&\hspace{20mm} 
+ \left[  - \frac 75  \left( 3x^2+3x+6 \right) + 2 \left(x^3-2x^2+2x +3 \right)   (2e^x -1) \right] e^x 
\biggr\}
\nonumber \\
&=&
\frac2{(1+e^x )^4} \left[
1 + \left(- x^3- \frac65 x^2+ \frac45 x -\frac{107}{5} \right) e^x + \left( 2x^3-4x+19 \right)e^{2x}  \right].
\label{ddrrff2}
\end{eqnarray}

For $-7/10< x\le0$,  we have $e^x \le 1$ and then $\left[ (r f)^2 \right]''$ is
bounded from above as
\begin{eqnarray}
\left[ (r f)^2 \right]'' &\le&
\frac2{(1+e^x )^4} \left[
1 + \left(- x^3- \frac65 x^2+ \frac45 x -\frac{107}{5} \right) e^x+ \left( 2x^3-4x+19 \right) e^x  \right]
\nonumber \\
&=&
\frac2{(1+e^x )^4} \left[
  1 + \left( x^3- \frac65 x^2- \frac{16}{5} x -\frac{12}{5} \right)  e^x \right] \nonumber \\
& <& \frac2{(1+e^x )^4} \ <\ 2.
\end{eqnarray}

For $0<x\le 1$, we have $e^x \le e <3$, and it gives
\begin{eqnarray}
\left[ (r f)^2 \right]'' &<&
\frac2{(1+e^x )^4} \left[
1 + \frac13 \left(- x^3- \frac65 x^2+ \frac45 x -\frac{107}{5} \right) e^{2x}+ \left( 2x^3-4x+19 \right)e^{2x} \right]
\nonumber \\
&=&
\frac2{(1+e^x )^4} + \frac{2e^{2x}}{15 (1+e^x )^4} \left[
 25 x^3 -6 x^2 -56 x + 178 \right] .
 \label{1032370}
\end{eqnarray}
The part in the square bracket of the second term 
in the second line is positive in the interval $0<x\leq 1$
since its function has
a minimal value at $x=(6+2\sqrt{1059})/75 ~(<1)$ and a maximal value at $x=(6-2\sqrt{1059})/75 ~(<0)$.
Then, the maximum value in $0<x\leq 1$ is under 178.
%
%
Moreover, we can estimate its coefficient as
\begin{eqnarray}
\frac{2e^{2x}}{15(1+e^x )^4}< \frac{1}{120},
\end{eqnarray}
where we used the fact that ${e^{2x}}/{(1+e^x )^4}$
monotonically decreases for $x>0$.

Then, $\left[ (r f)^2 \right]''$ is evaluated as 
\begin{eqnarray}
\left[ (r f)^2 \right]'' 
&< &
\frac2{(1+e^x )^4} + \dfrac{178}{120} 
\nonumber \\
&<&
\dfrac{1}{8} +\dfrac{178}{120} 
\ <\ 2
\label{0e}
\end{eqnarray}

\subsubsection{The case $e < e^x $~~({\it i.e.} $1<x$)}

We estimate 
the first and second lines of Eq.~\eq{ddrrff2}. 
The terms in the square bracket in the second line are shown to be positive as 
\begin{eqnarray}
&& - \left( x^3+x^2+5x+9 \right) + \left(2x^3-4x^2+4x +6 \right)  e^x
\nonumber \\
&& \hspace{10mm}
= - \left( x^3+x^2+5x+9 \right) + 2 \left[ x(x-1)^2 +x+3 \right]  e^x
\nonumber \\
&& \hspace{10mm}
> - \left( x^3+x^2+5x+9 \right) + 5 \left[ x(x-1)^2 +x+3 \right]
\nonumber \\
&& \hspace{10mm}
= 4x^3-11x^2+5x+6 >0,
\end{eqnarray}
where we used $5/2 < e < e^x $ and $x>1$ in the first inequality and used $x>1$ in the second inequality.
Since for $x>1$ we have
${\be}/{(r+\be)} < x+1$ [see Eq.~\eq{ineqre}], 
the value of $\left[ (r f)^2 \right]''$ is bounded as 
\begin{eqnarray}
\left[ (r f)^2 \right]''
&<&
\frac2{(1+e^x )^4} \left\{
1 + \left( x^3-x^2+9x -7 \right)e^x + \left( -2x^3+8x^2-12x+7 \right)e^{2x} \right.
 \nonumber \\ 
&&\hspace{20mm} 
\left.+ \left[  - \left( x^3+x^2+5x+9 \right) + \left(2x^3-4x^2+4x +6 \right) e^x \right]e^x (1+x)
\right\}
 \nonumber \\ 
&=&
\frac2{(1+e^x )^4} \left\{
1 - \left( x^4 + x^3+7x^2+5x +16 \right) e^x \right.
 \nonumber \\ 
&&\hspace{40mm} 
\left.+ \left( 2x^4-4x^3+8x^2-2x+13 \right)e^{2x} \right\}
\nonumber \\ 
&<&
\frac2{(1+e^x )^4}+ \frac{2 e^{2x}}{(1+e^x )^4} \left( 2x^4-4x^3+8x^2-2x+13 \right).
\label{ddrrff3}
\end{eqnarray}
Here, the first term has the maximum value $2 (1+e)^{-4} <1/50$ at $x=1$. 
The derivative of the second term is shown to be negative as
\begin{eqnarray}
&&\frac{d}{dx} \left[ \frac{2 e^{2x}}{(1+e^x )^4} \left( 2x^4-4x^3+8x^2-2x+13 \right)\right]
 \nonumber \\ 
&& \hspace{20mm}
= \frac{4 e^{2x}}{(1+e^x )^5} \left[ \left( 2x^4+2x^2+6x+12 \right)
- e^x \left(2x^4 -8 x^3 + 14 x^2 -10 x +14 \right) \right]
 \nonumber \\ 
&& \hspace{20mm}
<\frac{4 e^{2x}}{(1+e^x )^5} \biggl[ \left( 2x^4+2x^2+6x+12 \right)
 \nonumber \\ 
&& \hspace{50mm}
- \left( 1 + x + \frac12 x^2 \right) \left(2x^4 -8 x^3 + 14 x^2 -10 x +14 \right) \biggr]
 \nonumber \\ 
&& \hspace{20mm}
=-\frac{4 e^{2x}}{(1+e^x )^5} \left[ \left( x^3-x^2-x-1 \right)^2 + x(x-1)(x+4) +3x^2+1 \right]
 <0,
\end{eqnarray}
where we used 
\begin{eqnarray}
2x^4 -8 x^3 + 14 x^2 -10 x +14 = 2 (x-2)^2 x^2 + 5 (x-1)^2 + x^2 + 9> 0
\end{eqnarray}
and $e^x > 1 + x + x^2/2$.
Thus, the second term of the right-hand side of Eq.~\eq{ddrrff3}
is a decreasing function. 
The maximum of the right-hand side of the inequality
of Eq.~\eqref{ddrrff3} occurs at $x=1$ and its value satisfies 
\begin{eqnarray}
\frac{34 e^2}{(1+e)^4} < \frac{14}{10}.
\end{eqnarray}
Therefore, the value of $\left[ (r f)^2\right]''$ is less than 2.


\end{document}